 \documentclass[aps,prc,twocolumn,showpacs,preprintnumbers,amsmath,amssymb,a4paper]{revtex4}

\usepackage{amssymb}
\usepackage{graphics}
\usepackage{psfrag}
\usepackage{epsfig}
\usepackage{bm}
\def\bea{\begin{eqnarray}}
\def\eea{\end{eqnarray}}
\def\beq{\begin{equation}}
\def\eeq{\end{equation}}

\newcommand{\ppbar}{{\bar p p}}
\newcommand{\nnbar}{{\bar N  N}}
\newcommand{\lbarl}{{\bar \Lambda \Lambda}}
\newcommand{\lcbarlc}{\bar{\Lambda}_c^- {\Lambda}_c^+}
\newcommand{\ddbar}{D\bar{D}}
\newcommand{\dsds}{{D_s^+}{D_s^-}}
\newcommand{\kkbar}{\bar{K}{K}}
\begin{document}
\title{Production of charmed pseudoscalar mesons in antiproton-proton annihilation}
\author{J. Haidenbauer$^1$ and G. Krein$^2$}

\affiliation{
$^1$Institute for Advanced Simulation and J\"ulich Center for Hadron
Physics, Forschungszentrum J\"ulich, D-52425 J\"ulich, Germany \\
$^2$Instituto de F\'{\i}sica Te\'{o}rica, Universidade Estadual
Paulista, Rua Dr. Bento Teobaldo Ferraz, 271 - Bloco II, 01140-070 S\~{a}o Paulo, SP, Brazil
}

\begin{abstract}
We study the production of charmed mesons ($D$, $D_s$) 
in antiproton-proton ($\bar pp$) annihilation close 
to the reaction thresholds. The elementary charm production 
process is described by baryon exchange and in the constituent quark model,
respectively. Effects of the interactions in the
initial and final states are taken into account rigorously. 
The calculations are performed in close analogy to our earlier
study on $\bar pp \to \bar KK$ by connecting the processes 
via SU(4) flavor symmetry. 
Our predictions for the $\ddbar$ production cross section are in the 
order of $10^{-2}$ -- $10^{-1}$ $\mu b$. They turned out to be
comparable to those obtained in other studies. 
The cross section for a $\dsds$ pair is found to be of the same
order of magnitude despite the fact that its production in $\bar pp$ 
scattering requires a two-step process. 
\end{abstract}


%

\maketitle

\section{Introduction}
\label{intro}
Physics involving charmed particles is one of the main topics to be
explored at the planned FAIR facility in Darmstadt \cite{PANDA,Hohne}. 
In particular, the program proposed by the $\bar {\rm P}$ANDA Collaboration 
encompasses a wide range of activities connected to this subject including 
high-accuracy spectroscopy of charmed hadrons and the investigation of their 
interactions with ordinary matter \cite{Wiedner}. 
Presently very little is known about the interaction of charmed particles with
conventional hadrons and/or nuclear matter built up predominantly from up- and 
down quarks. Clearly, the rate at which charmed hadrons can be produced is a crucial 
factor for designing and performing secondary experiments with those particles. 
In particular, attaining a sufficient yield is a prerequisite for investigating issues 
like $c\bar{c}$-quarkonium dissociation~\cite{Matsui:1986dk} and the creation of
new e\-xo\-tic nuclear bound states of $J/\psi$ and $\eta_c$~\cite{{Brodsky:1989jd},
{Ko:2000jx},{Krein:2010vp},{Tsushima:2011kh},{Krein:2013rha}}, charmed 
hypernuclei~\cite{hyper-c}, and charmed D-mesic nuclei~\cite{{Tsushima:1998ru},
{GarciaRecio:2010vt},{GarciaRecio:2011xt}} that have been discussed in the
literature over the last few years. 

In this work we present predictions for the charm-production reactions 
$\ppbar \rightarrow \ddbar$ and $\ppbar \rightarrow \dsds$ close to their 
thresholds. The work builds on the J\"ulich meson-baryon model for the 
reaction $\ppbar \rightarrow \kkbar$~\cite{Hippchen1,Hippchen2,Mull}. 
The extension of the model from the strangeness to the charm
sector follows a strategy similar to our recent work on the $DN$ 
and ${\bar D}N$ interactions~\cite{Hai071,Hai072,Hai11}, and
on the reaction $\ppbar \rightarrow \lcbarlc$~\cite{HK10},
namely by imposing as a working hypothesis SU(4) symmetry constraints
and improvements from quark-gluon dynamics at short distances~\cite{{Hadjimichef:2000en},
{Hadjimichef:2002xe}}. 
The microscopic charm-production process is described by baryon 
exchange ($\Lambda_c$, $\Sigma_c$) and the transition potentials are 
derived from the corresponding transitions in the strangeness-production 
channels ($\bar KK$) utilizing values of the involved coupling constants
that are fixed from SU(4) symmetry.
The reaction amplitudes themselves are evaluated in distorted-wave 
Born approximation (DWBA). This is done because we want to take into account 
rigorously the effects of the initial ($\ppbar$) and also of the final-state 
interactions which are known to play an important role for energies near the 
production threshold \cite{Haiden1991,Mull91,Haiden1992,KohnoL,Alberg,KohnoM}.

As before in our study of the reaction $\ppbar \rightarrow \lcbarlc$~\cite{HK10}
we investigate the effect of replacing the transition interaction based
on meson-baryon dynamics by a charm-production potential derived in the constituent
quark model. 
This allows us to shed light on the model dependence of our results. Furthermore, 
we compare our predictions with the ones of other model calculations of the 
$\ppbar \to \ddbar$ reaction from the literature
\cite{Kroll:1988cd,Kaidalov:1994mda,Kerbikov,Titov:2008yf,Mannel12,Goritschnig13}. 
In some of those studies a quark-gluon description based on 
a factorization hypothesis of hard and soft processes \cite{Kroll:1988cd,Goritschnig13}
is employed, while in others a non-perturbative quark-gluon string model is used, based 
on secondary Regge pole exchanges including absorptive corrections~\cite{Kaidalov:1994mda,
Titov:2008yf,Mannel12}. 
Preliminary results (for $\ppbar \to \ddbar$) of our study were presented in \cite{HK11}.

The paper is organized as follows: In Sect.~\ref{sec:2} we discuss the
$\nnbar$ interaction used for the initial-state interaction (ISI). 
Because of the known sensitivity of the annhilation cross sections on 
the ISI, we examine its effect by considering various $\nnbar$ potentials
where we make sure that all of them reproduce the total $\ppbar$ cross section 
in the relevant energy range and, in general, describe also data on
integrated elastic and charge-exchange cross sections and even
$\ppbar$ differential cross sections. 
Predictions for the reaction $\ppbar \to \ddbar$ that include effects of
the $\ppbar$ ISI are presented in Sect.~\ref{sec:3}.
Transition potentials based on meson-baryon dynamics and derived in the 
quark model are considered. 
 
To study also the influence of the final-state interaction (FSI),
we extend a $\pi\pi-\kkbar$ interaction potential, developed by the 
J\"ulich group in the past \cite{Lohse90,Janssen95}, 
by adding to this coupled-channel model the $\ddbar$ and 
the $\dsds$ channels. This extension is described in detail in 
Sect.~\ref{sec:4} and then the influence of the resulting FSI on the
$\ppbar \to \ddbar$ cross section is examined. Finally, we provide
predictions for the production of the charmed strange meson $D_s$ 
in the reaction $\ppbar \to \dsds$. 
In this case a two-step process is required and, therefore, the pertinent
cross section cannot be calculated in approaches that rely on the 
Born approximation in one way or the other. However, in a coupled-channel
approach like ours the transition amplitude from $\ppbar$ to $\dsds$ is 
generated in a natural way.    
The paper ends with a summary. 
Technical aspects related to the derivation of the various interaction
potentials are summarized in the Appendices. 

\section{Interaction in the initial $\bar NN$ system}
\label{sec:2}

For the $\bar NN$ interaction in the initial state, we take the
same model that has been already employed in our recent study of 
the process $\ppbar \rightarrow \lcbarlc$ \cite{HK10}. 
It is based on the interaction originally developed for our
investigation of $\ppbar \to \lbarl$
and consists of an elastic part which is deduced (via G-parity transform)
from a simple, energy-independent one-boson-exchange $NN$ potential
(OBEPF) \cite{obepf} and a phenomenological annihilation part for 
which a spin-, isospin-, and energy-independent optical potential of 
Gaussian form is adopted: 
\beq
V^{\nnbar \to \nnbar}_{opt}(r)  = (U_0 + i W_0) \, e^{- r^2/2r^2_0} .
\label{Vaa-pp}
\eeq

\begin{table}[ht]
\renewcommand{\arraystretch}{1.1}
\centering
\caption{\label{cross} Total and integrated elastic $\ppbar$ cross
sections and integrated charge-exchange ($\ppbar\to \bar nn$) cross sections 
in $mb$ for the four potentials considered in comparison to experimental values. 
}
\vskip 0.2cm
\begin{tabular}{|c|l|ccccr|}
\hline
& $p_{lab}$ (GeV/c) & \ A \ & \ A' \ & \ B \ & \ C \ & Experiment \\
\hline
$\sigma_{tot}$ 
&6.65 & $56.6$ & $59.1$ & $57.0$ & $56.9$ & $59.5\pm 0.5$ \cite{Denisov}\\
&7.30 & $56.0$ & $58.5$ & $56.3$ & $56.3$ & $58.3\pm 1.3$ \cite{Patel}\\
&9.10 & $54.7$ & $56.9$ & $54.7$ & $54.8$ & $57.51\pm 0.73$ \cite{Gregory}\\
&10.0 & $54.2$ & $56.4$ & $54.1$ & $54.3$ & $54.7\pm 0.60$ \cite{Denisov}\\
&12.0 & $53.5$ & $55.6$ & $53.3$ & $53.4$ & $51.7\pm 0.80$ \cite{Johnson}\\
\hline
\hline
$\sigma_{el}$ 
&6.0 & $15.9$ & $15.1$ & $16.7$ & $15.9$ & $15.6\pm 0.8$ \cite{Ambats}\\
&7.2 & $15.2$ & $14.3$ & $15.8$ & $15.2$ & $13.79\pm 1.0$ \cite{Foley}\\
&8.0 & $14.9$ & $14.0$ & $15.4$ & $14.8$ & $12.88\pm 0.1$ \cite{Russ}\\
&8.9 & $14.6$ & $13.6$ & $15.0$ & $14.5$ & $13.89\pm 0.35$ \cite{Foley}\\
&10.0 & $14.4$ & $13.4$ & $14.6$ & $14.2$ & $14.6\pm 3.3$ \cite{Foley}\\
&12.0 & $14.0$ & $13.0$ & $14.1$ & $13.8$ & $11.59\pm 0.41$ \cite{Foley}\\
&     &        &        &        &        & $11.34\pm 0.6$ \cite{Johnson}\\
\hline
\hline
$\sigma_{cex}$ 
&6.0 & $0.50$ & $0.57$ & $0.55$ & $0.78$ & $0.563\pm 0.082$ \cite{Astbury}\\
&7.0 & $0.38$ & $0.45$ & $0.42$ & $0.64$ & $0.373\pm 0.054$ \cite{Astbury}\\
&7.76 & $0.32$ & $0.37$ & $0.36$ & $0.57$ & $0.380\pm 0.042$ \cite{Lee}\\
&9.0 & $0.24$ & $0.29$ & $0.28$ & $0.47$ & $0.284\pm 0.041$ \cite{Astbury}\\
\hline
\end{tabular}
\end{table}
\renewcommand{\arraystretch}{1.0}

The parameters of the potential ($U_0$, $W_0$, $r_0$) can be found
in Ref.~\cite{HK10}. They were determined by a fit to $\nnbar$ data in 
the energy range relevant for the reaction $\ppbar \to \lcbarlc$, 
specifically to total cross sections \cite{Denisov,Foley,Astbury}
around $p_{lab} =$ 10 GeV/c, i.e. close to the $\lcbarlc$
threshold which is at 10.162 GeV/c. A comparison of the model
results with the data on total and integrated elastic and
charge-exchange cross sections but also with differential 
$\ppbar$ cross sections \cite{Foley,Berglund} around 10 GeV/c
was presented in Ref.~\cite{HK10}. 

The thresholds for the reactions $\ppbar \rightarrow \ddbar$
and $\ppbar \rightarrow \dsds$ are at somewhat lower momenta,
namely, at 6.442 and 7.255~GeV/c, respectively. Therefore,
we present here again $\nnbar$ results, but now we compare them 
with experiments over a momentum range that covers also the 
thresholds of those two reactions we study in the present work. 
Integrated cross sections for the considered $\nnbar$ interactions
are summarized in Table~\ref{cross}.

\begin{figure*}[t]
\begin{center}
\includegraphics[height=75mm,angle=-90]{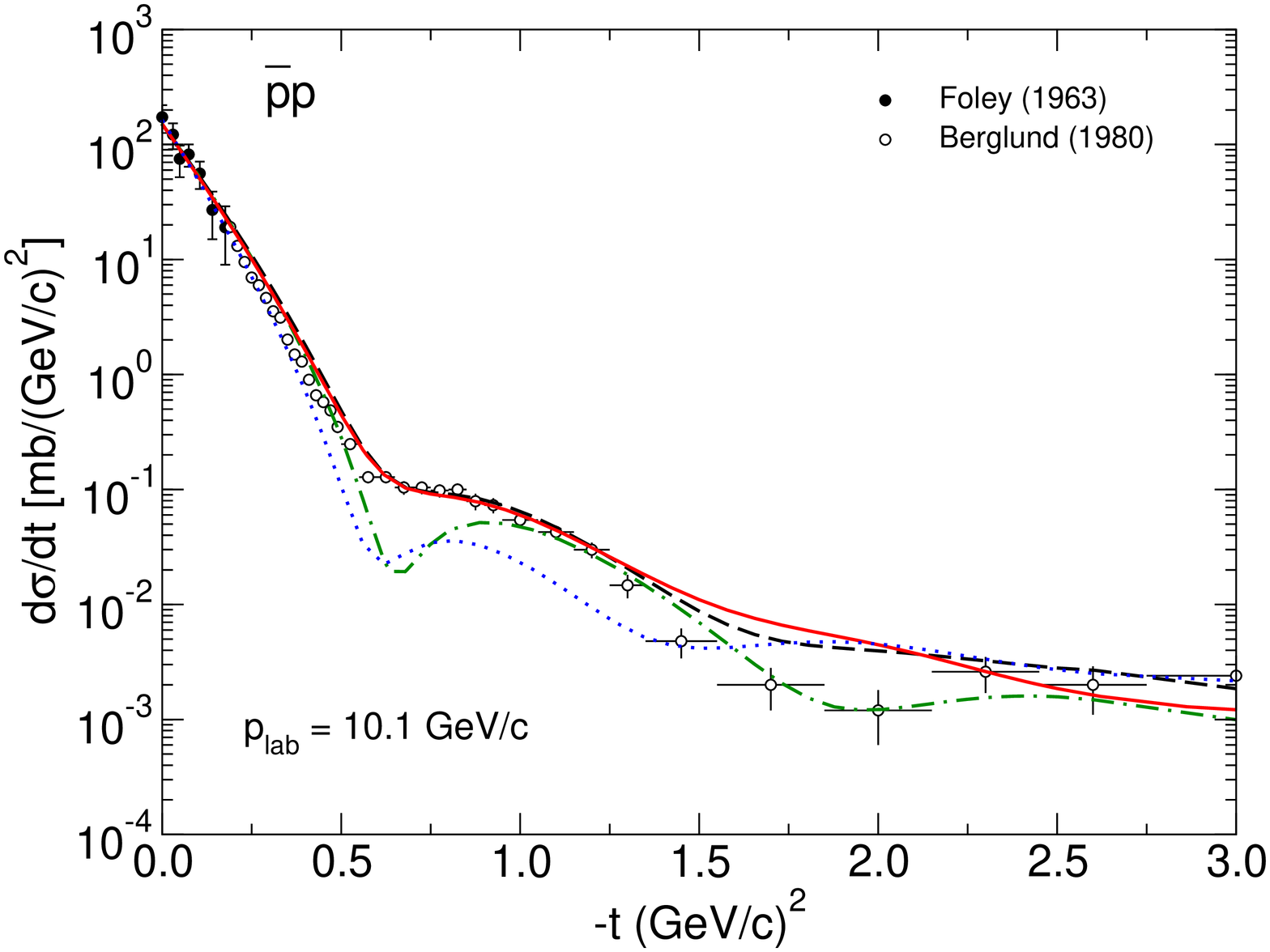}
\includegraphics[height=75mm,angle=-90]{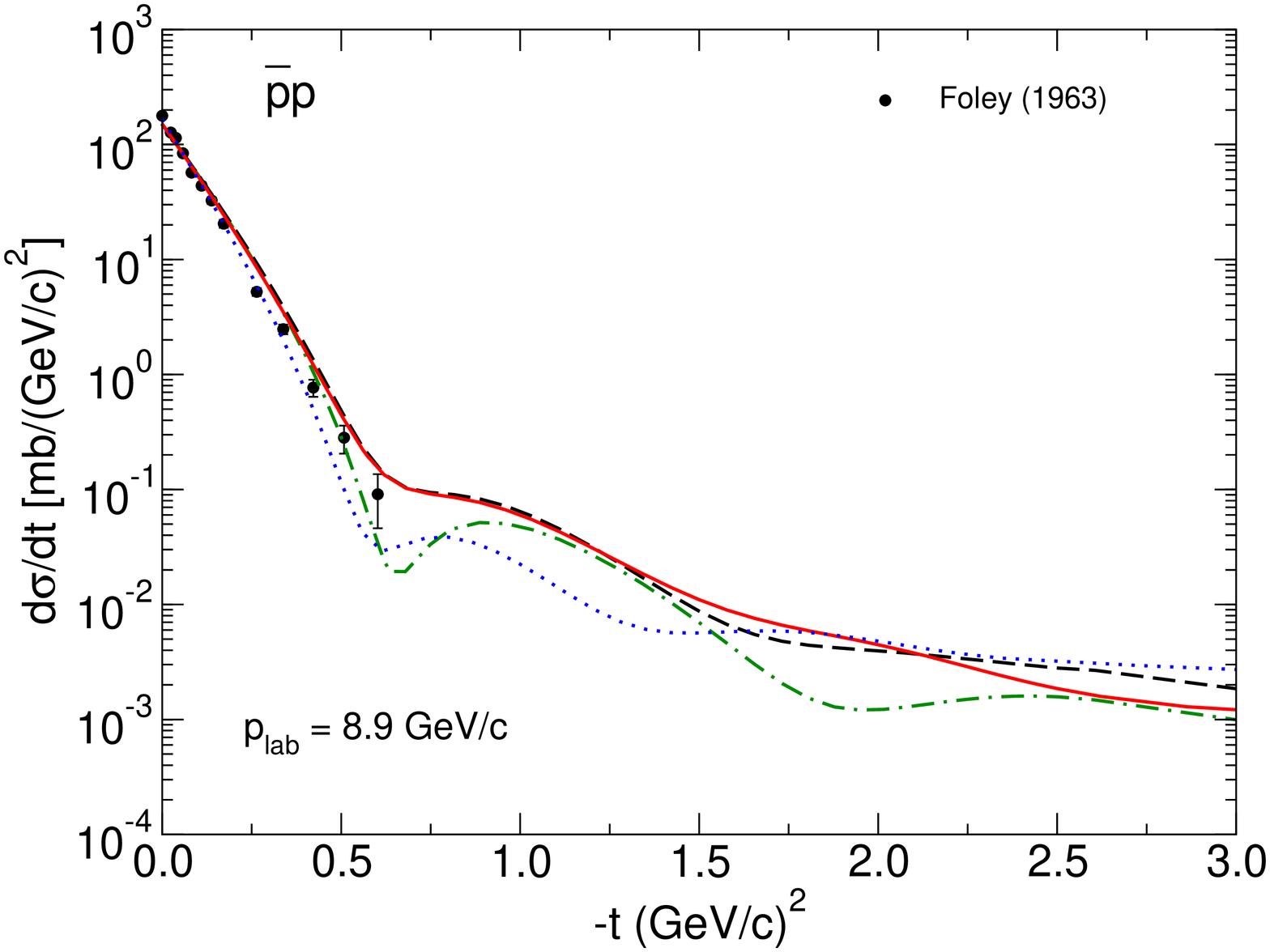}
\includegraphics[height=75mm,angle=-90]{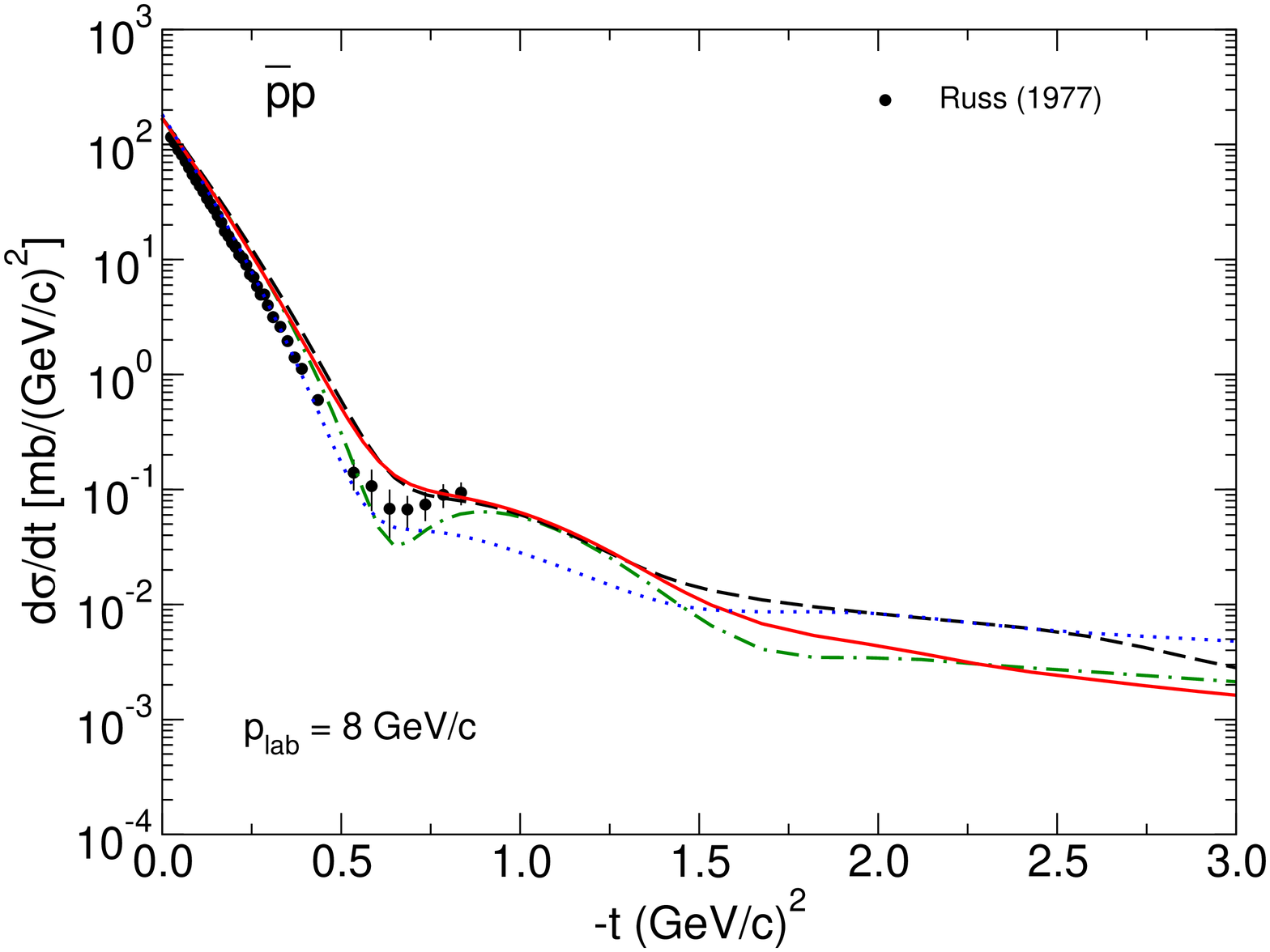}
\includegraphics[height=75mm,angle=-90]{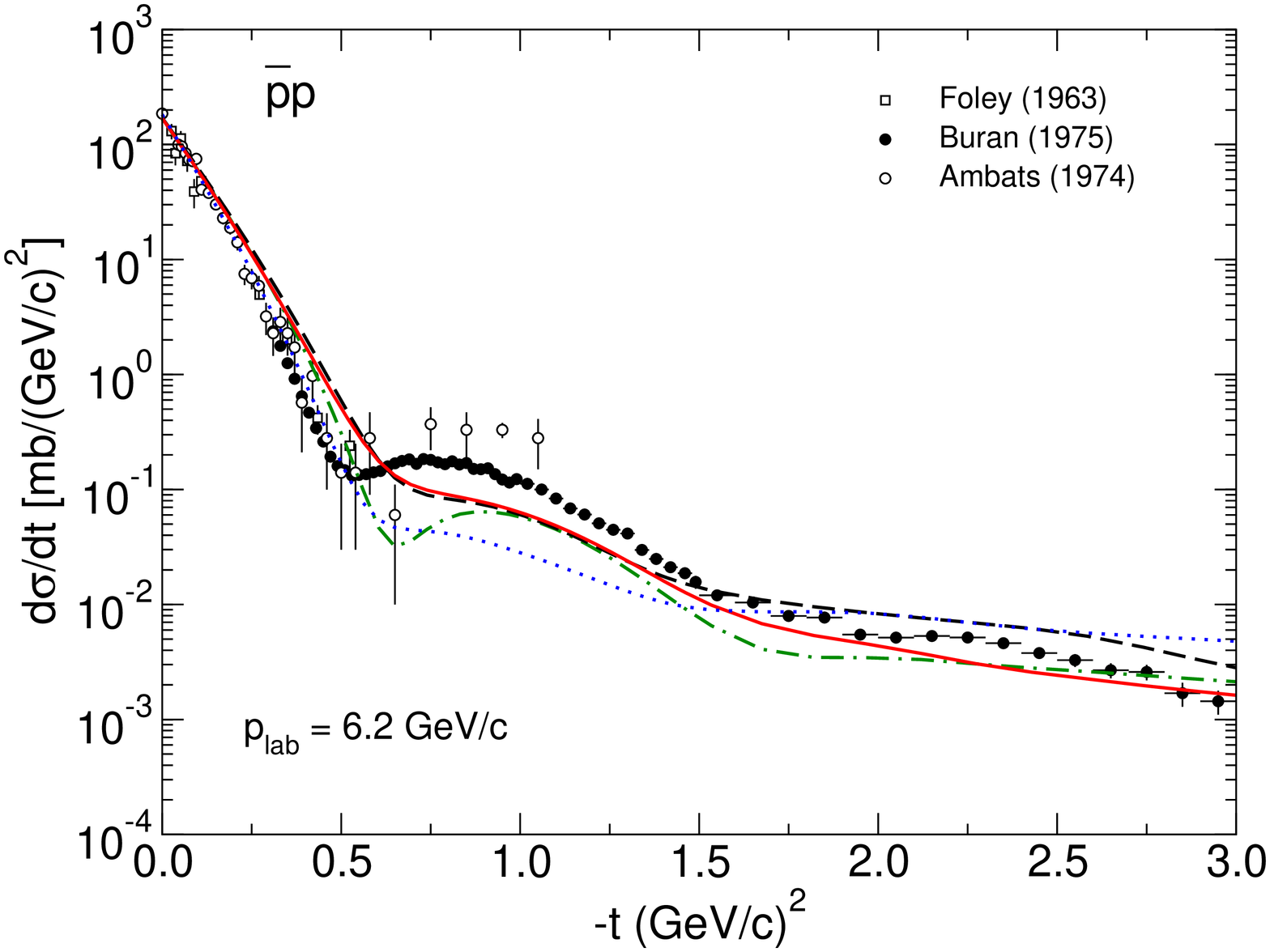}
\caption{Differential cross section for elastic $\ppbar$ scattering at
$p_{lab}$ = 6.2, 8, 8.9, and 10.1 GeV/c as a function of $t$. 
The dash-dotted curve corresponds
to a calculation where only one-pion exchange is added to the optical
potential (A). The dashed and solid curves are obtained by leaving out
vector-meson exchanges (B) or by reducing the elastic part (except for
the pion exchange) to 10~\% (C), respectively. The dotted
line is an alternative fit, made to reproduce specifically the slope 
of the data at 6.2 GeV/c, where likewise only pion exchange is added
to the optical potential (A').
The experimental 
information is taken from Foley et al. \cite{Foley}, Berglund et 
al. \cite{Berglund}, Russ et al. \cite{Russ},
Buran et al. \cite{Buran}, and Ambats et al. \cite{Ambats}.
}
\label{ppbar}
\end{center}
\end{figure*}

As already noted in Ref.~\cite{HK10}, at the high energies where $D$ and $D_s$
production occurs any $NN$ potential has to be considered as being purely 
phenomenological and, therefore, one has to question whether fixing the 
elastic part of the $\nnbar$ potential via G-parity by utilizing such an $NN$ 
interaction that was fitted to low-energy $NN$ data is still meaningful.  
In addition, one knows from studies on $\ppbar \to \lbarl$ that the 
magnitude of the cross sections depends very sensitively on the 
ISI \cite{Haiden1991,Haiden1992,KohnoL,Alberg}.
Specifically, the absorptive character of the $\nnbar$ interaction
leads to a strong reduction of the cross section as compared to
results obtained in the Born approximation, i.e., based on the transition
potential alone. 
Because of these reasons, in Ref.~\cite{HK10} several variants of the $\nnbar$ 
model were considered which differed in the treatment of the elastic
part, with the intention to use them for illustrating the uncertainties 
in the predictions due to the used $\ppbar$ interaction. 
In all those $\nnbar$ interaction potentials (denoted by A, B, C, and D in 
Ref.~\cite{HK10}), the longest ranged (and model-independent) part of the
elastic $\ppbar$ interaction, namely one-pion exchange, was kept,
but the shorter ranged contributions, consisting of vector-meson 
and scalar-meson exchanges, were treated differently. In the
present investigation we do not consider model D which does not provide
a realistic description of the $\nnbar$ data in the energy region
relevant for $D$ and $D_s$ production. 
Instead we consider a new fit, called $A'$, that includes only one-pion exchange 
for the elastic part (like $A$) but yields a better reproduction of the
somewhat stronger fall off of the differential cross section 
exhibited by the data around 6.2 GeV/c; see Fig.~\ref{ppbar}.
In any case, in all scenarios a rather satisfying description of the $\nnbar$ 
data in the region 6--10 GeV/c is obtained; cf. Table \ref{cross} and Fig.~\ref{ppbar}. 
In particular, not only the slope but in some cases even the shoulder in the 
differential cross section is reproduced quantitatively by these interactions.

\section{The reaction $\ppbar \rightarrow \ddbar$}
\label{sec:3}

\subsection{$\ppbar \rightarrow \ddbar$ based on baryon exchange}

\begin{figure}[t]
\includegraphics[height=180mm]{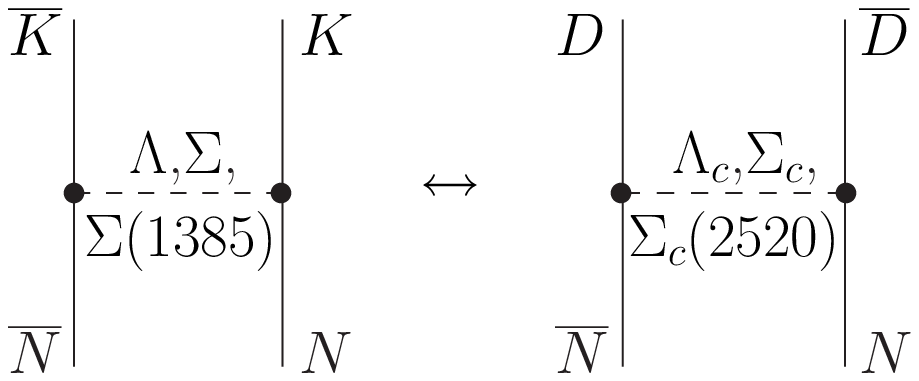}
\vskip -14.0cm 
\caption{Transition potential for 
$\nnbar \to \ddbar$ (right) and $\nnbar \to \kkbar$ (left),
respectively. 
}
\label{Diadd}
\end{figure}

\begin{figure}[hb]
\includegraphics[height=75mm,angle=-90]{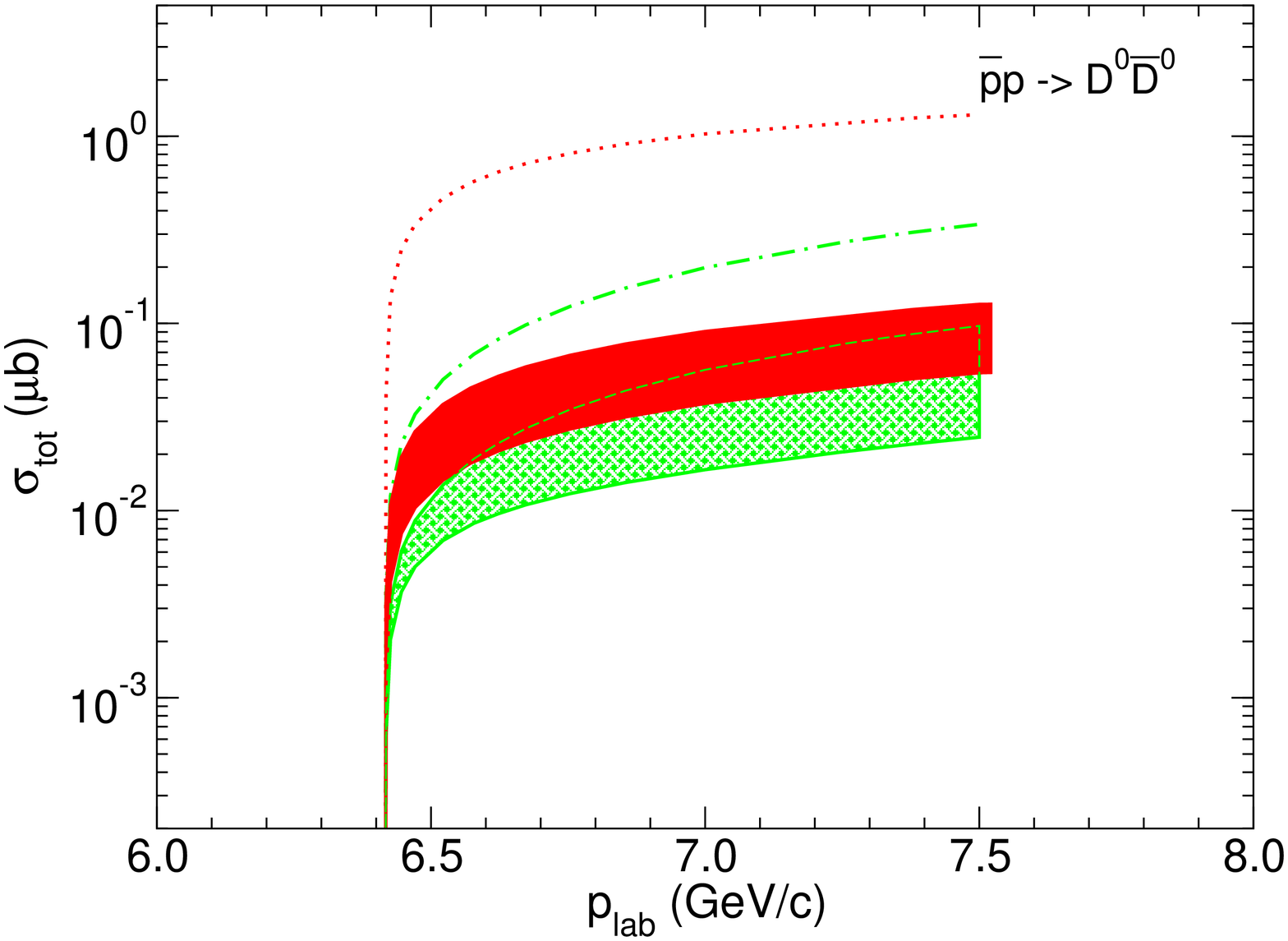}
\includegraphics[height=75mm,angle=-90]{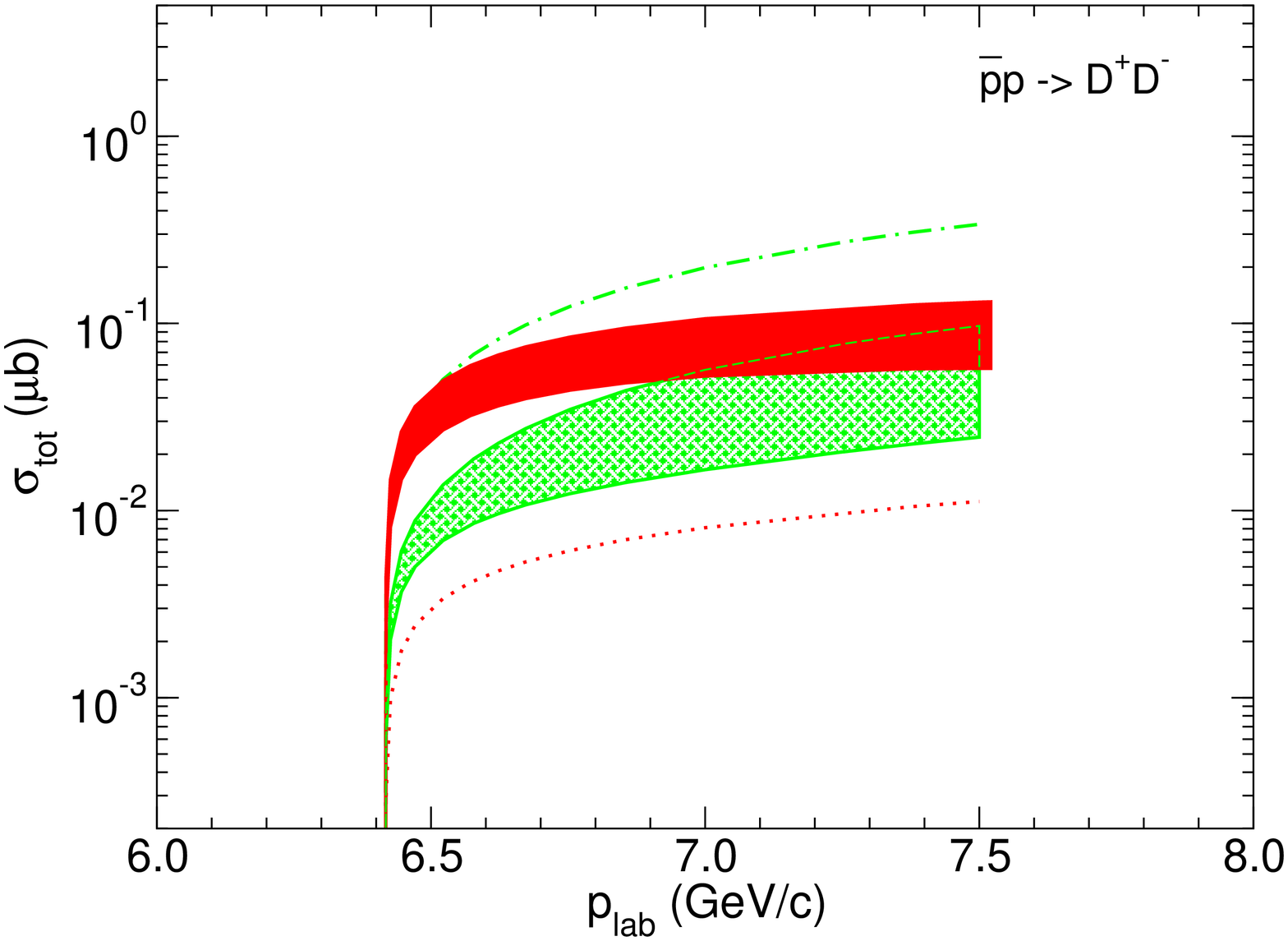}
\caption{Total reaction cross sections for
$\ppbar \to \ddbar$ as a function of $p_{lab}$,
based on baryon exchange (shaded band) and the quark model (grid).
Results obtained in Born approximation are indicated by the dotted 
(baryon exchange) and dash-dotted (quark model) lines, respectively. 
}
\label{fig:dd} 
\end{figure}

Within meson-baryon dynamics, 
the transition from $\ppbar$ to $\ddbar$ is generated by the
exchange of charmed baryons, in particular the $\Lambda_c$ and
$\Sigma_c$ (in analogy to the exchange of $\Lambda$ and $\Sigma$
in case of the reaction $\ppbar \to \kkbar$); see Fig.~\ref{Diadd}. 
Explicit expressions for the transition potentials
can be found in Appendix A of Ref.~\cite{Hippchen1}. 
They are of the generic form
\beq
V^{\nnbar \rightarrow \ddbar} (t)
\sim \sum_{Y=\Lambda_c^+,\Sigma_c,\Sigma_c^*} 
\frac{ f^2_{YND} \, F^2_{YND}(t)}{\omega_D(\sqrt{s}-E_N-\omega_D -E_Y)} ,
\label{Vtrans}
\eeq
where $f_{YND}$ are coupling constants, $F_{YN D}(t)$
are vertex form factors, and $E_N$, $\omega_D$, $E_Y$ are
the energies of the nucleon, $D$-meson and the exchanged
baryon, respectively. Under the assumption of $SU(4)$ symmetry 
the pertinent coupling constants are given by 
\begin{eqnarray}
f_{\Lambda_c^+ N D} &=& 
-\frac{1}{\sqrt{3}}(1+2{\alpha}) f_{NN\pi} \approx {-1.04}\, f_{NN\pi}, 
\nonumber \\
f_{\Sigma_c N D} &=&
(1-2{\alpha}) f_{NN\pi} \approx {0.2}\, f_{NN\pi}, 
\label{SiCou}
\end{eqnarray}
where we assumed for the F/(F+D) ratio ${\alpha} \approx 0.4$. 
Thus, one expects that $\Lambda_c^+$ exchange dominates the transition
while $\Sigma_c$ exchange should be suppressed. Specifically, 
the isospin decomposition 
\begin{eqnarray}
\nonumber
{V}^{\ppbar \to {D^0 \bar D^0}} &=&
\frac{1}{2}(V_{I=0}^{\nnbar \to \ddbar}+V_{I=1}^{\nnbar \to \ddbar}) , \\ 
{V}^{\ppbar \to {D^+ D^-}} &=&
\frac{1}{2}(V_{I=0}^{\nnbar \to \ddbar}-V_{I=1}^{\nnbar \to \ddbar}) , 
\label{Iso}
\end{eqnarray}
suggests that ${V}^{\ppbar \to {D^0 \bar D^0}} \gg {V}^{\ppbar \to {D^+ D^-}}$ 
because the (dominant) contribution of the isoscalar $\Lambda_c^+$ exchange 
drops out in the latter channel. 
Indeed, within the Born approximation, the cross sections predicted 
for $D^0 \bar D^0$ are more than two orders of magnitude larger than
those for $D^+ D^-$, cf. the dotted lines Fig.~\ref{fig:dd}. 
(The coupling constant $f_{N\Sigma^* K}$, and accordingly for
$f_{N\Sigma_c^*  D}$, is likewise very small \cite{Holzenkamp:1989tq} so 
that the contribution of $\Sigma^*$ ($\Sigma_c^*$) exchange turns out 
to be negligible.) 

The vertex form factors adopted in Refs.~\cite{Hippchen1,Hippchen2}
for the $\nnbar$ annihilation diagrams are not of the conventional
monopole type but involve fourth powers of the cutoff mass $\Lambda$,
of the exchanged baryon, and of the transferred momentum, see
Eq.~(2.15) in Ref.~\cite{Hippchen1}. Such a more complicated parameterization
was required in order to avoid unphysical singularities in the potential.
We employ the same form here. In the actual calculation a cutoff mass 
$\Lambda$ of 3.5 GeV at the $YND$ vertices is used. 
This choice is motivated by the experience gained in our studies of
$\nnbar \to MM$ annihilation processes in the past and, specifically,
in $\nnbar \to \kkbar$ where cutoff masses that are roughly
1 GeV larger than the masses of the exchanged baryons were found 
to be appropriate. 
We will come back to (and explore) the sensitivity of the
results to variations of the cutoff mass below.

Let us now focus on the effects of the initial state interaction. 
Those effects are included by solving the formal coupled-channel
equations
\begin{eqnarray}
\nonumber
T^{\nnbar,\nnbar} &=& V^{\nnbar,\nnbar} \\
&+& V^{\nnbar,\nnbar} G^{\nnbar} T^{\nnbar,\nnbar} \ , \\
\nonumber
T^{\ddbar,\nnbar} &=& V^{\ddbar,\nnbar} \\
&+& V^{\ddbar,\nnbar} G^{\nnbar} T^{\nnbar,\nnbar} \ , 
\label{DWBA}
\end{eqnarray}
utilizing the $\nnbar$ potential described in Sect.~\ref{sec:2}. 
Of~course, Eq.~(\ref{DWBA}) implies that the $\nnbar \to \ddbar$ transition
amplitude is effectively evaluated in a DWBA. 

Results with the inclusion of ISI effects are presented as bands
in Fig.~\ref{fig:dd} because we consider several variants of the $\nnbar$ 
potential as discussed in the previous section.
It is obvious that the results change drastically once the ISI is included 
in the calculation. The cross sections for $D^0 \bar D^0$ are strongly
reduced while at the same time those for $D^+ D^-$ are enhanced. 
Indeed now both $\ddbar$ channels are produced at a comparable rate. 
In fact, the predicted cross section for $D^+ D^-$ appears to be even 
somewhat larger than the one for $D^0 \bar D^0$. 

Whereas the reduction in the $D^0 \bar D^0$ case is in line with
comparable effects observed in the previous studies of $\nnbar$ annihilation 
processes \cite{Haiden1991,Haiden1992,KohnoL,Alberg}, as mentioned above, 
the enhancement seen for $D^+ D^-$ may be somewhat surprising, at least
at first sight. 
However, it can be easily understood if one recalls that
the $\Lambda_c^+$ cannot contribute to the $\ppbar \to D^- D^+$ transition 
potential as discussed above. Only $\Sigma_c$ (and 
$\Sigma_c^*$) exchange contributes. But their coupling constants are very
small according to SU(4) symmetry (cf. Eq.~(\ref{SiCou})) and the somewhat 
larger masses reduce the importance of $\Sigma_c$-exchange contributions further. 
This is the reason why the $\ppbar \to D^+ D^-$ cross section is strongly
suppressed in Born approximation. 
The consideration of the ISI via the
employed DWBA approach (\ref{DWBA}) generates two-step transitions of the 
form $\ppbar \to \bar nn \to D^+ D^-$. In this case $\Lambda_c^+$ 
exchange is no longer absent because it does contribute to the
$\bar nn \to D^+ D^-$ transition potential and, accordingly, those 
two-step transitions are enhanced in comparison to the Born approximation.

\begin{figure}[h]
\includegraphics[scale=0.65]{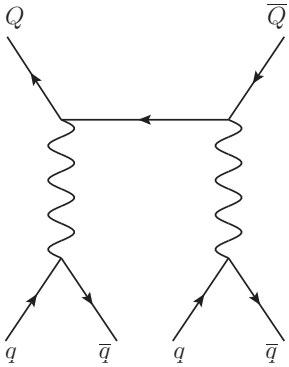}
\caption{Microscopic quark-model mechanism for the transition potential: 
annihilation of two pairs of light quarks, $q\bar{q}=u\bar{u},d\bar{d}$,
and creation of a pair of heavier quarks, $Q\bar{Q}=s\bar{s},c\bar{c}$.
}
\label{fig:qqQQ}
\end{figure}

\subsection{$\ppbar \rightarrow \ddbar$ based on the quark model}

We consider a $\ppbar \to \ddbar$ transition potential derived in a constituent 
quark model where two light quark pairs ($\bar{u}u$ and $\bar{d}d$) are 
annihilated and a charmed quark pair ($\bar{c}$c) is created -- see Fig.~\ref{fig:qqQQ}. 
We base our study on the model of Kohno and Weise~\cite{KohnoM} for the 
$\ppbar \to \kkbar$ reaction; we replace parameters corresponding to 
the $s-$quark and $K-$meson of that model by those of the $c-$quark and
$D-$meson. The quark-model $\nnbar \to \ddbar$ transition potential 
$V^{\nnbar\rightarrow \ddbar}_Q(t)$ can be written as
\begin{equation}
V^{\nnbar\rightarrow \ddbar}_Q(t) = \chi^{\dag}_{\bar N} \left[ 
h_1(t) \, {\bm \sigma}\cdot{\bf p} + h_2(t) \, {\bm \sigma}\cdot{\bf p}'
\right] \chi_N ,
\label{VppQM}
\end{equation}
where ${\bf p}$ and ${\bf p}'$ are the $\nnbar$ and $\ddbar$ center-of-mass 
(c.m.) momenta, $\chi_N$ and $\chi_{\bar N}$ are the spin Pauli spinors of the nucleon
and antinucleon, and $h_1(t)$ and $h_2(t)$ depend upon quark masses and hadron
sizes, and the effective strength of quark-pair annihilation and creation  
-- their explicit expressions are given in Appendix~\ref{app:QM}. A specific
feature of the quark-model potential is that $V^{\ppbar\rightarrow D^0\bar{D}^0}_Q 
= - V^{\ppbar\rightarrow D^+{D}^-}_Q$ (see Appendix~\ref{app:QM}), so that 
there is no isospin $I=0$ transition. This is in contrast to the transitions 
induced by $\Lambda_c^+$ and $\Sigma_c$ exchange, as discussed above. 

\begin{figure}[ht]
\includegraphics[height=75mm,angle=-90]{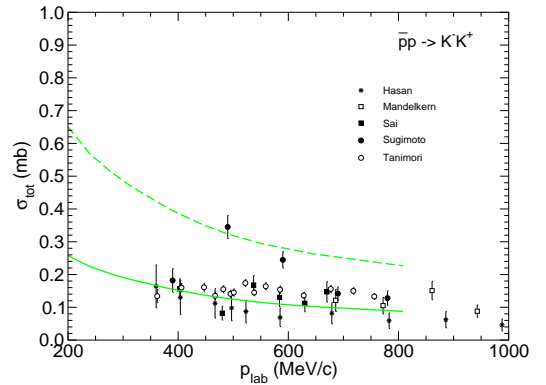}
\caption{Cross section for $\ppbar \to K^-K^+$ scattering as a function of $p_{lab}$. 
Results are based on the quark model. The curves 
correspond to different values for the effective coupling strength $\alpha_A/m^2_G$ 
-- 0.12 fm$^2$ (solid line) or 0.15 fm$^2$ (dashed line) -- see discussion in the text. 
Data are taken from Refs.~\cite{Mandelkern,Sai,Tanimori,Sugimoto,Hasan}.}
\label{KKbcr}    
\end{figure}

Before presenting the results for $\ppbar\rightarrow \ddbar$, let us first examine 
the performance of the model in the reaction $\ppbar\rightarrow K^-K^+$ for which
there are experimental data available. We use standard quark-model values for quark 
masses and size parameters (they are given in Appendix~\ref{app:QM}). 
And to facilite a comparison with the results of Kohno and Weise
we use the same value for the effective coupling strength $\alpha_A/m^2_G$ as 
in their study of that reaction, namely $\alpha_A/m^2_G=0.15$~fm$^2$. 
The employed ISI is the same as for the $\ddbar$ case discussed above, but
with parameters of the optical potential fitted to low-energy $\nnbar$ data 
(cf. OBEPF in Table IV of Ref.~\cite{obennb}). 
As visible from Fig.~\ref{KKbcr} (dashed line) the result is roughly in line with the 
available data and it is also close to the original result of Kohno and Weise \cite{KohnoM}. 
The differences are presumably due to the different ISI used by them and by us. 
Actually, with a slight reduction of the effective coupling strength 
($\alpha_A/m^2_G=0.12$~fm$^2$), the bulk of the $K^-K^+$ data can be quantitatively
reproduced; see the solid curve in the figure. Thus, we will use this smaller
coupling constant in the following calculations of charmed meson production to be on 
the safe side. 

The quark model results for $\ppbar \rightarrow \ddbar$ are shown in Fig.~\ref{fig:dd}. 
Clearly, because the transitions $D^+{D}^-$ and $D^0\bar{D}^0$ are of the same 
magnitude, the corresponding cross sections calculated in Born approximation 
are the same. Moreover, for the same reason, the two-step transitions 
$\ppbar \to \bar NN \to D^+ D^-$ and $\ppbar \to \bar NN \to D^0 \bar{D}^0$ that
make up the ISI provide equal reductions for both final states. Figure~\ref{fig:dd}
also reveals that the quark model and baryon-exchange transitions yield 
comparable predictions, with those of the quark model being on average smaller 
by a factor roughly equal to~3. In addition, the results show once more the fundamental 
role played by the ISI in the $\ppbar$ annihilation process, as the two transition 
mechanisms have very different isospin dependence and yet the final results are
of comparable magnitude.

\begin{figure*}
\includegraphics[height=75mm,angle=-90]{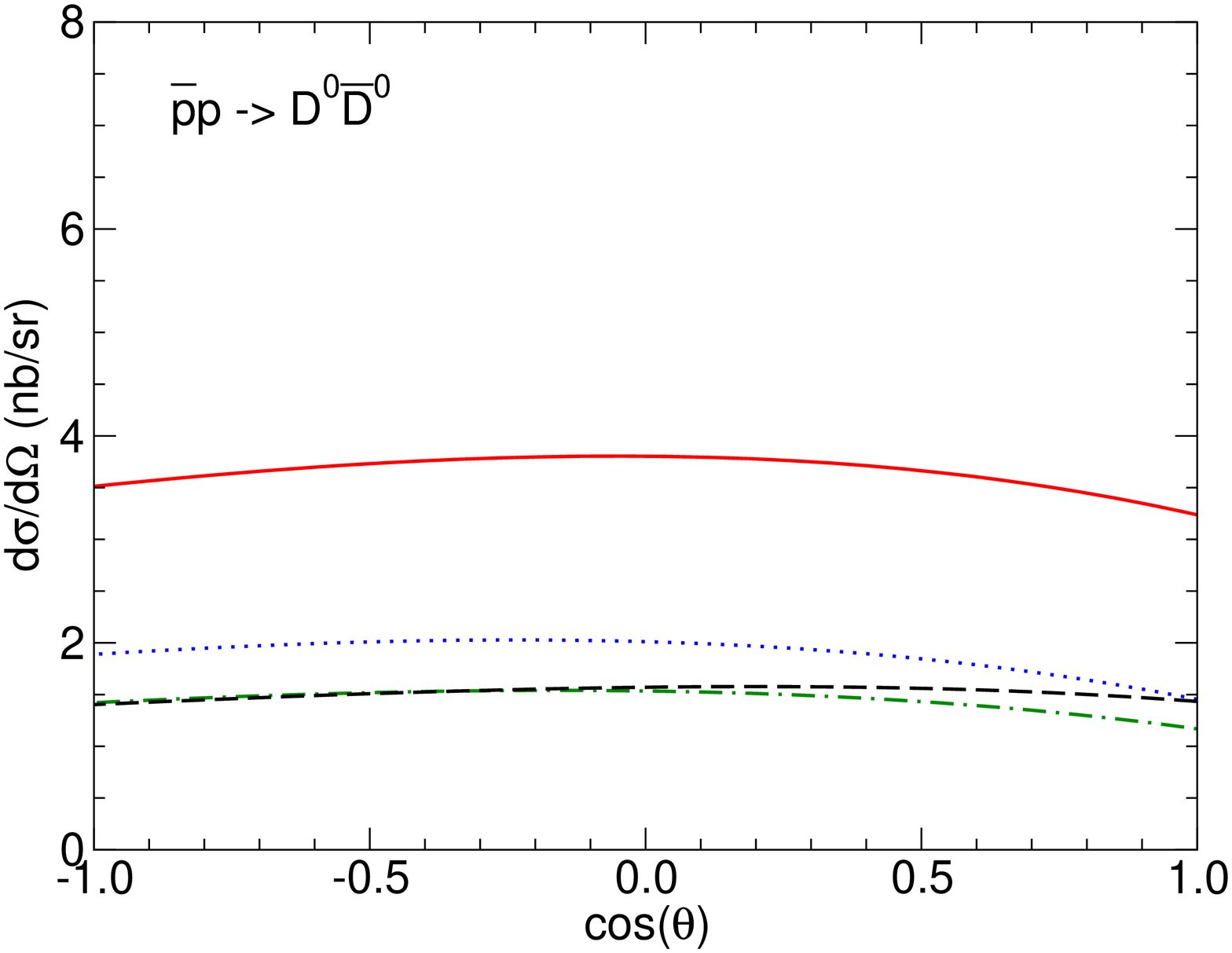}
\includegraphics[height=75mm,angle=-90]{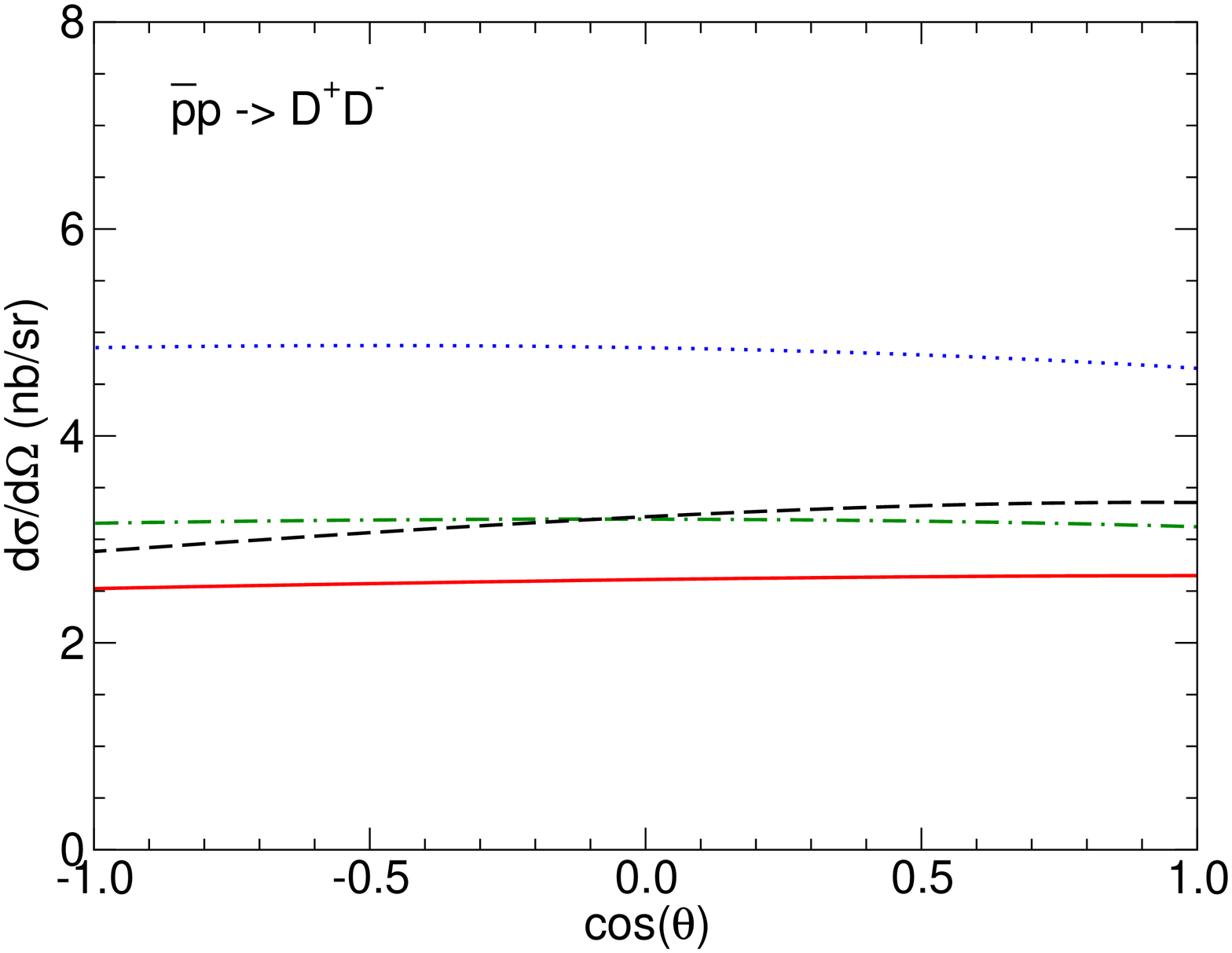}
\caption{Differential cross sections for
$\ppbar \to \ddbar$ at $p_{lab} = 6.578$ GeV/c (excess energy $\epsilon = 40$ MeV).
Results for different ISI are shown, namely for model
A (dash-dotted curve), A' (dotted curve), B (dashed curve), and C 
(solid curve).  
}
\label{fig:4}       
\end{figure*}

Predictions for the differential cross sections based on the baryon-exchange 
transition potential are presented in Fig.~\ref{fig:4} at the excess 
energy $\epsilon = 40$ MeV (corresponding to $p_{lab} = 6.578$ GeV/c). 
We show the results for the different ISI separately so that one can see the 
variations induced by the individual $\nnbar$ potentials. The overall variation at 
this energy amounts to roughly a factor~2. In all cases there is only a
rather weak dependence of the $D^0\bar D^0$ and $D^+D^-$ cross sections
on the scattering angle which is a clear sign for the dominance of 
$s$-wave production. This is not surprising in view of the fact that
the production mechanism is of rather short range. 
In this context it is instructive to recall the selection rules for the 
production of two pseudoscalar mesons \cite{Hippchen1}. Conservation of
total angular momentum and parity implies that the lowest two 
partial-wave amplitudes are given by the transitions $^3P_0 \to s$
and $^3S_1 \to p$ where the first symbol characterizes the $\nnbar$ partial
wave in the standard spectral notation and the second specifies the angular
momentum in the $\ddbar$ (or $\kkbar$) system. Dominance of the $s$-wave 
is therefore expected near the $\ddbar$ threshold. 
However, in the case of $\ppbar \rightarrow \kkbar$ one is actually close to 
the $\ppbar$ threshold so that the $\nnbar$ system is in the $^3S_1$ partial 
wave and the $\kkbar$ system will be dominantly produced in a $p$ wave. 
Indeed, for that reaction, one observes a pronounced angular dependence of the 
differential cross section already at moderate energies, in the experimental 
data but also in model calculations \cite{Hippchen2}.

The differential cross sections for $\ppbar \to \ddbar$ based on the constituent 
quark model exhibit a very similar behavior and, therefore, we refrain from showing them.  

Finally, let us mention that reducing the cutoff mass $\Lambda$ from 3.5 to 3 GeV 
in the baryon-exchange transition potential reduces the cross section by roughly a 
factor~5. Thus, the cutoff dependence appears to be somewhat stronger here than 
what we observed for $\ppbar \to \lcbarlc$ where the cross section dropped by a
factor of around 3 for a comparable variation of the cutoff mass \cite{HK10}.
The variation of the cutoff mass simulates to some extent a possible SU(4)
breaking in the $YND$ coupling constants because, like a direct variation of 
the coupling constants, it changes the strength of the potential in the
relevant (physical) region of the momentum transfer $t$. 
Indeed, results for the
$\Lambda_c ND$ and $\Sigma_c ND$ coupling constants from QCD sum rules
\cite{Mannel12} suggest a moderate breaking of SU(4) symmetry. Interestingly,
the coupling constants for charmed baryons turned out to be somewhat larger 
than their strange counterparts which, naively seen, would imply larger cross sections.
In particular, the reported breaking of the SU(4) symmetry of $1.47^{+0.58}_{-0.44}$ 
in terms of the ratio of the $\Lambda_c ND$ to $\Lambda NK$ coupling constants \cite{Mannel12},
amounts to roughly a factor~5 on the level of the cross sections for the central
value. Unfortunately, the theoretical uncertainty for the ratio is large, so that, in 
principle, its value is even compatible with $1$, i.e. with the SU(4) result. 
In any case, it is worthwhile to note that the variation in the cross sections deduced
from the SU(4) breaking in the coupling constants is of very similar magnitude as
the one suggested by our variation of the cutoff mass. 
In this context let us say that only a very small deviation from SU(4) symmetry, i.e. 
in the order of $1.05$ in terms of the ratio of the $\Lambda_c ND$ to $\Lambda NK$ 
coupling constants, is obtained within the $^3P_0$ constituent quark 
model~\cite{Krein:2012lra}.

The comparison between the results based on baryon exchange and on the quark
model provides an alternative picture for the uncertainty in the $\ddbar$
production cross section, independent from the issue of SU(4) symmetry breaking.
Also here we see variations in the order of a factor 3-5, as mentioned above. 

\subsection{Comparison with other results}

In the literature one can find several other studies of the reaction 
$\ppbar \to \ddbar$.
The most recent publication is
by Goritschnig et al.~\cite{Goritschnig13}, who employ a
quark-gluon description based on a factorization hypothesis of hard and soft
processes. This work supersedes an earlier study by that group within
a quark-diquark picture, where already concrete predictions for the
$D^+D^-$ production cross section were given~\cite{Kroll:1988cd}.
In the study by Kaidalov and Volkovitsky~\cite{Kaidalov:1994mda}
a non-perturbative quark-gluon string model was used, based on secondary Regge pole
exchanges including absorptive corrections. On the same lines, there is the
more recent publication by Titov and K\"ampfer~\cite{Titov:2008yf}.
Finally, in the work by Khodjamirian et al.~\cite{Mannel12} the quark-gluon string 
model of Ref.~\cite{Kaidalov:1994mda} was revisited, but now strong coupling constants
calculated from QCD lightcone sum rules were employed. 

Interestingly, and in contrast to studies of the reaction $\ppbar \to \lcbarlc$ 
\cite{HK10}, the majority of the calculations for $\ppbar \to \ddbar$ predict 
cross sections that are pretty much of comparable magnitude, at least on 
a qualitative level. This is to some extent surprising because, as far
as we can see, none of the other studies take into account effects of the 
ISI which strongly influences the magnitude of our results. Of course, one
could argue that such effects are included effectively in the coupling
constants or the di-quark form factor say, employed in those other studies. 
Anyway, the results presented in \cite{Titov:2008yf} as well as those in
\cite{Mannel12} exhibit a strong suppression of the $D^+D^-$ cross section
as compared $D^0\bar D^0$ -- which as we argued above is definitely a 
consequence of the Born approximation together with $YND$ coupling constants 
that fulfill (approximate) SU(4) symmetry. 

On the quantitative level we see that the 
$D^0\bar D^0$ cross section of Kaidalov \cite{Kaidalov:1994mda} lies 
within the band of our results as provided in Fig.~\ref{fig:dd} and 
the same is true for the $D^+D^-$ cross section of \cite{Kroll:1988cd}
(in the energy region considered by us). The $D^0\bar D^0$ predictions
in \cite{Mannel12} are also comparable to ours but their
$D^+D^-$ cross sections are down by two orders of magnitude. 
The $\ddbar$ cross section of Kerbikov \cite{Kerbikov} and the
corrected $D^0\bar D^0$ cross section 
by Goritschnig et al.~\cite{Goritschnig13} (cf. the erratum) are about
one order of magnitude smaller. 
In Ref. \cite{Titov:2008yf} only 
differential cross sections are given. Because of that we calculated
${d \sigma / d}t$ at the excess energy {$\epsilon$ = 0.5 GeV}
and $t_{max} - t$ = 0.2 GeV$^2$ in order to facilitate a comparison. 
Our results for 
$\ppbar \to D^+ D^-$ and $\ppbar \to D^0\bar D^0$
are $(0.8-1.8) \times 10^{-2}$ and $(0.6-1.3) \times 10^{-2}$ 
$\mu b$/GeV$^2$, respectively, which should be compared to 
$\approx 1.5 \times 10^{-2}$ and $\approx 50 \times 10^{-2}$ $\mu b$/GeV$^2$
by Titov and K\"ampfer, estimated from their figures.

\section{Effects of the final $\ddbar$ interaction and 
the reaction $\ppbar \to \dsds$}
\label{sec:4}

In the study of the reaction $\ppbar \to \kkbar$ by the
J\"ulich group \cite{Mull} the interaction in the $\kkbar$ 
channel was ignored. Indeed, since the mass of the kaon
is significantly smaller than the one of the proton, already
at the $\ppbar$ threshold the relative momentum in the 
produced $\kkbar$ system is fairly large and,
therefore, one can expect that FSI effects are small in this
case. Moreover, as pointed out in Sect.~\ref{sec:3}.B, 
the $\kkbar$ pair is produced primarily in a $p$ wave near 
the $\ppbar$ threshold because of the section rules. 
Obviously, for $\ppbar \to \ddbar$ these arguments no longer 
hold! Thus, in the following we want to investigate, at least
qualitatively, the effect due to a FSI in the $\ddbar$ system, 
and we do this by adapting and
extending a $\pi\pi - \kkbar$ (coupled channels) model developed 
by the J\"ulich group some time ago \cite{Lohse90,Janssen95}. 

In the extension of the model we include not only the 
$\ddbar$ channel but also the $\dsds$ system. The mass of the
charmed strange meson $D_s$ is with 1969 MeV only about 100 MeV
larger than the one of the $D$ meson.
Thus, the thresholds of those two channels are relatively close to
each other, i.e. much closer than those of $\ddbar$ and $\kkbar$,
say, which could be of relevance for the $\ddbar$ FSI effects. 
In addition, and more interestingly, 
taking into account the $\dsds$ system enables us to provide also 
predictions for the reaction $\ppbar \to \dsds$ because then annihilation
into this channel becomes possible too, e.g., via the two-step process 
$\ppbar \to \ddbar \to \dsds$. 
The direct $\ppbar \to \dsds$ transition requires the annihilation of
three (up or down) quark-antiquark pairs and a creation of two
($s$ and $c$) quark-antiquark pairs and is, therefore, OZI suppressed. 
 
\begin{figure}[t]
\vspace*{+1mm}
\centerline{\hspace*{3mm}
\psfig{file=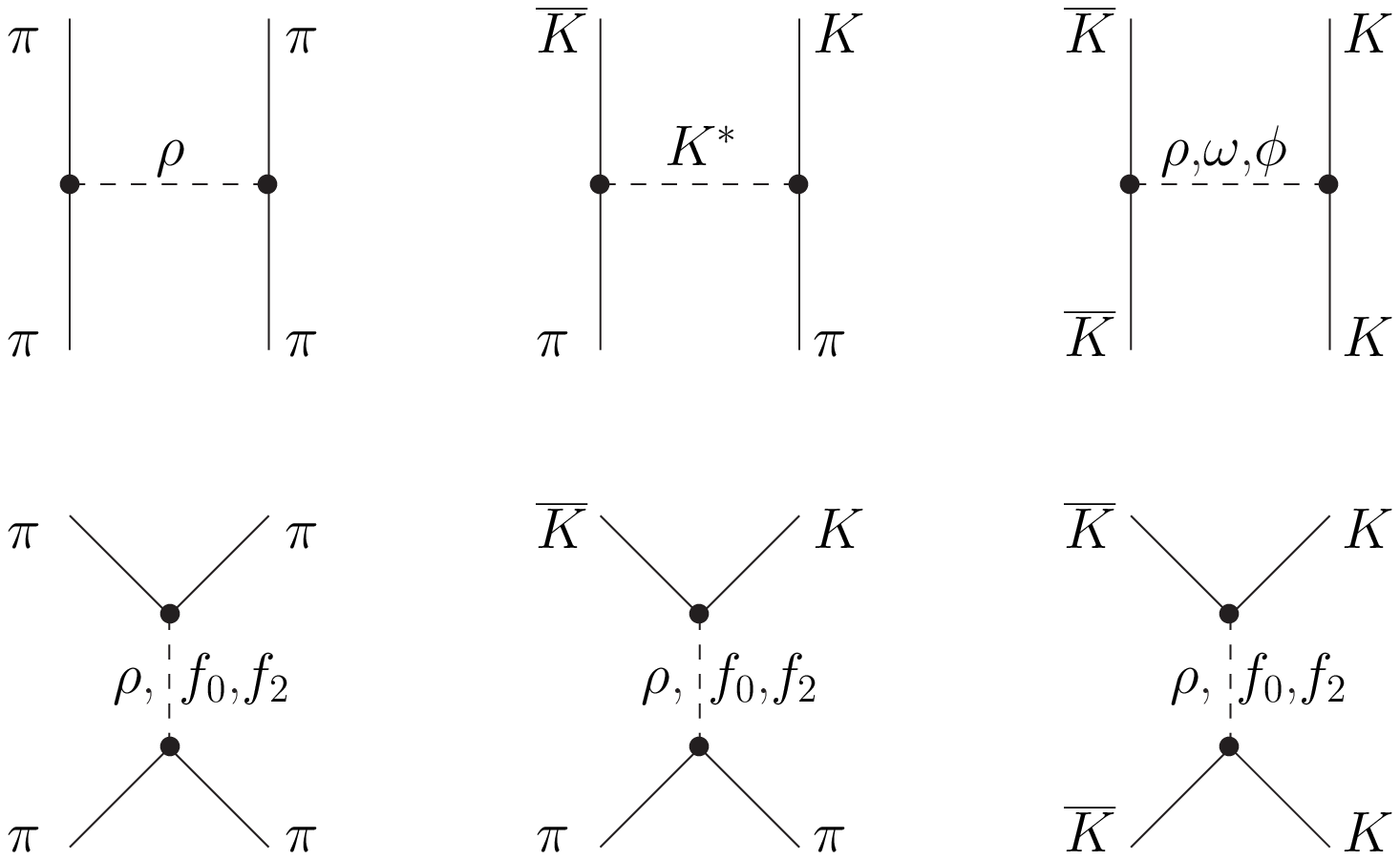,width=11.0cm,height=13.0cm}}
\vspace*{-7.5cm}
\caption{Diagrams included in the J\"ulich 
$\pi\pi- \kkbar$ potential~\cite{Janssen95}. 
}
\label{Diag1}
\end{figure}

The interactions in the $\ddbar$ and $\dsds$ systems are constructed 
along the lines of the J\"ulich meson exchange model for the $\pi\pi$ 
interaction for which the evaluation has been discussed in detail in Refs. 
\cite{Lohse90,Janssen95}. The present interaction is based 
on the version described in the latter reference. 
The potentials for $\pi\pi\rightarrow\pi\pi$, $\pi\pi\rightarrow
K\overline{K}$ and $K\overline{K} \rightarrow K \overline{K}$ are
generated from the diagrams shown in Fig.~\ref{Diag1}. The figure
contains only $s$- and $t$-channel diagrams; $u$-channel processes
corresponding to the considered $t$-channel processes are also included
whenever they contribute. The scalar-isoscalar particle denoted by
$\epsilon$ in Fig.~\ref{Diag1} effectively includes the singlet
and the octet member of the scalar nonet.  
The effects of $t$-channel $f_2(1270)$ and $\epsilon$ exchange were 
found to be negligible \cite{Janssen95} and, therefore, not
included in the model. 

The coupling constant $g_{\rho\pi\pi}$, required for $t$- and
$u$-channel exchange diagrams, is determined from the decay widths of
the $\rho$. Most of the other coupling constants are determined from 
SU(3) symmetry relations, and standard assumptions
about the octet/singlet mixing angles, as demonstrated in Ref.\
\cite{Lohse90}.

The scattering amplitudes are obtained by iterating these potentials
by using a coupled channel scattering equation, formally given by   
\begin{eqnarray}
T^{i,j}=&& V^{i,j} + \sum_l V^{i,l} G^l T^{l,j} 
\label{scaeq}
\end{eqnarray}
with $i,j,l=\pi\pi, \ \pi\eta, \ K\overline{K}$. 

This interaction yields a good description of the $\pi\pi$
phase shifts up to partial waves with total angular momentum
$J=2$ and for energies up to $\sqrt{s}\approx$ 1.4 GeV as
can be seen in Ref.~\cite{Janssen95}. Furthermore, as a special
feature, the $f_0(980)$ meson results as a dynamically generated 
state, namely as a quasi-bound $\bar KK$ state. 
Also the $a_0(980)$ is found to be dynamically generated in
the corresponding $\pi\eta-\bar KK$ system. 

\begin{figure}[t]
\vspace*{+1mm}
\centerline{\hspace*{3mm}
\psfig{file=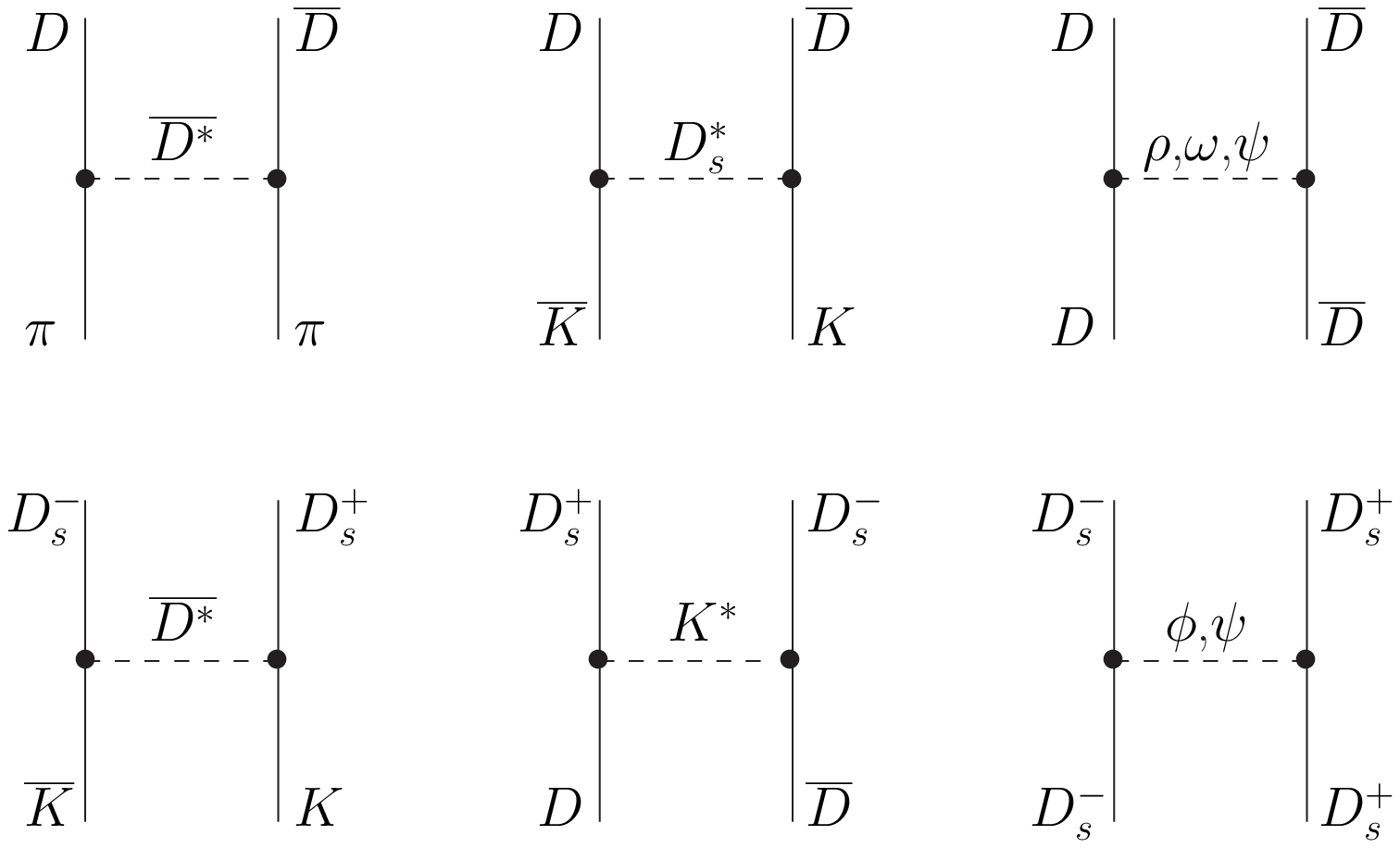,width=11.0cm,height=13.0cm}}
\vspace*{-7.5cm}
\caption{Additional diagrams that contribute to the potential 
when the $\ddbar$ and $D_s^-D_s^+$ channels are included. 
}
\label{Diag2}
\end{figure}

The additional diagrams that arise for the direct $\ddbar$ 
and $\dsds$
potentials and for the transitions from $\pi\pi$ and/or
$\kkbar$ to those channels are displayed in Fig.~\ref{Diag2}. 
In this extension we are again guided by SU(4) symmetry. 
Thus, we include $t$-channel exchanges of those vector mesons 
which are from the same SU(4) multiplet as those included
in the original J\"ulich model and, moreover, we assume that
all coupling constants at the additional three-meson vertices 
are given by SU(4) relations. The latter are summarized in
Appendix~\ref{app:su4}. 
As can be seen in Fig.~\ref{Diag1} the original J\"ulich
model includes also $s$-channel (resonance) diagrams, 
specifically, $\pi\pi/K\overline{K}\rightarrow \epsilon,\rho,f_2
\rightarrow\pi\pi/K\overline{K}$, which enable a unified 
description of all partial waves \cite{Janssen95}. 
However, those resonances lie far below the thresholds of the 
$\ddbar$ and $\dsds$ channels.
Therefore, they have very little influence on the results for
$\ddbar$ and $\dsds$ scattering as we verified in test 
calculations where we assumed that the bare coupling constants 
of those resonances to $\ddbar$ are the same as those for 
$\kkbar$. Thus, in the present extension of the model 
\cite{Janssen95} to the charm sector we set their couplings to 
the $\ddbar$ system to zero.

\begin{figure}
\includegraphics[height=75mm,angle=-90]{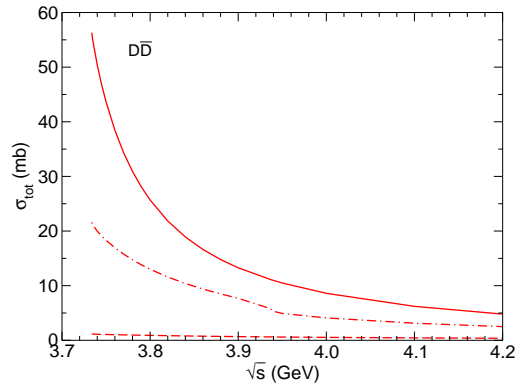}
\caption{Cross section for $\ddbar$ scattering in the $s$ wave
as a function of $\sqrt{s}$. The solid line is the results for 
the isospin $I=0$ channel and the dashed line for the $I=1$ channel. 
The dash-dotted curve indicates the changes in the $I=0$ case
when the coupling to $\bar D_sD_s$ is included in the model. 
}
\label{DDbcr}    
\end{figure}

Since the $\ddbar$ interaction was considered before and,
specifically, in a meson-exchange approach \cite{Liu10}
we display here also some prediction of the present
model. Cross sections for $\ddbar$ scattering in the
isospin $I=0$ and $I=1$ states can be found in Fig.~\ref{DDbcr}.
The main difference in the dynamics between our model and the 
one in Ref.~\cite{Liu10} is that the latter includes also
the exchange of scalar mesons. As mentioned above, $t$-channel exchange 
of a scalar meson has been considered in the original J\"ulich 
$\pi\pi-\kkbar$ potential \cite{Janssen95} but was found to be 
negligible. Because of that we neglected contributions from
scalar meson also in our extension to the charm sector. 

\begin{figure}[htb]
\includegraphics[height=75mm,angle=-90]{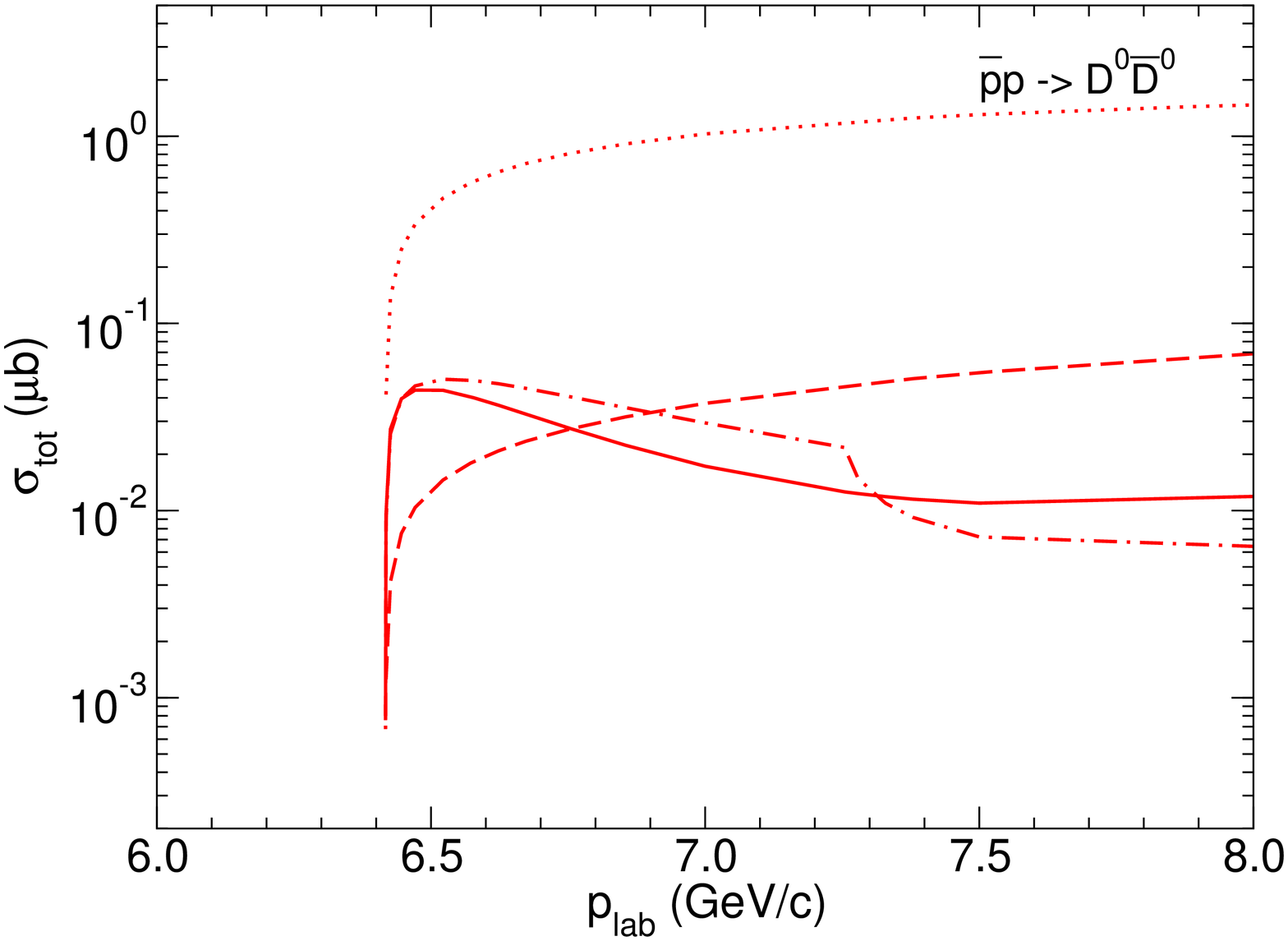}\\
\includegraphics[height=75mm,angle=-90]{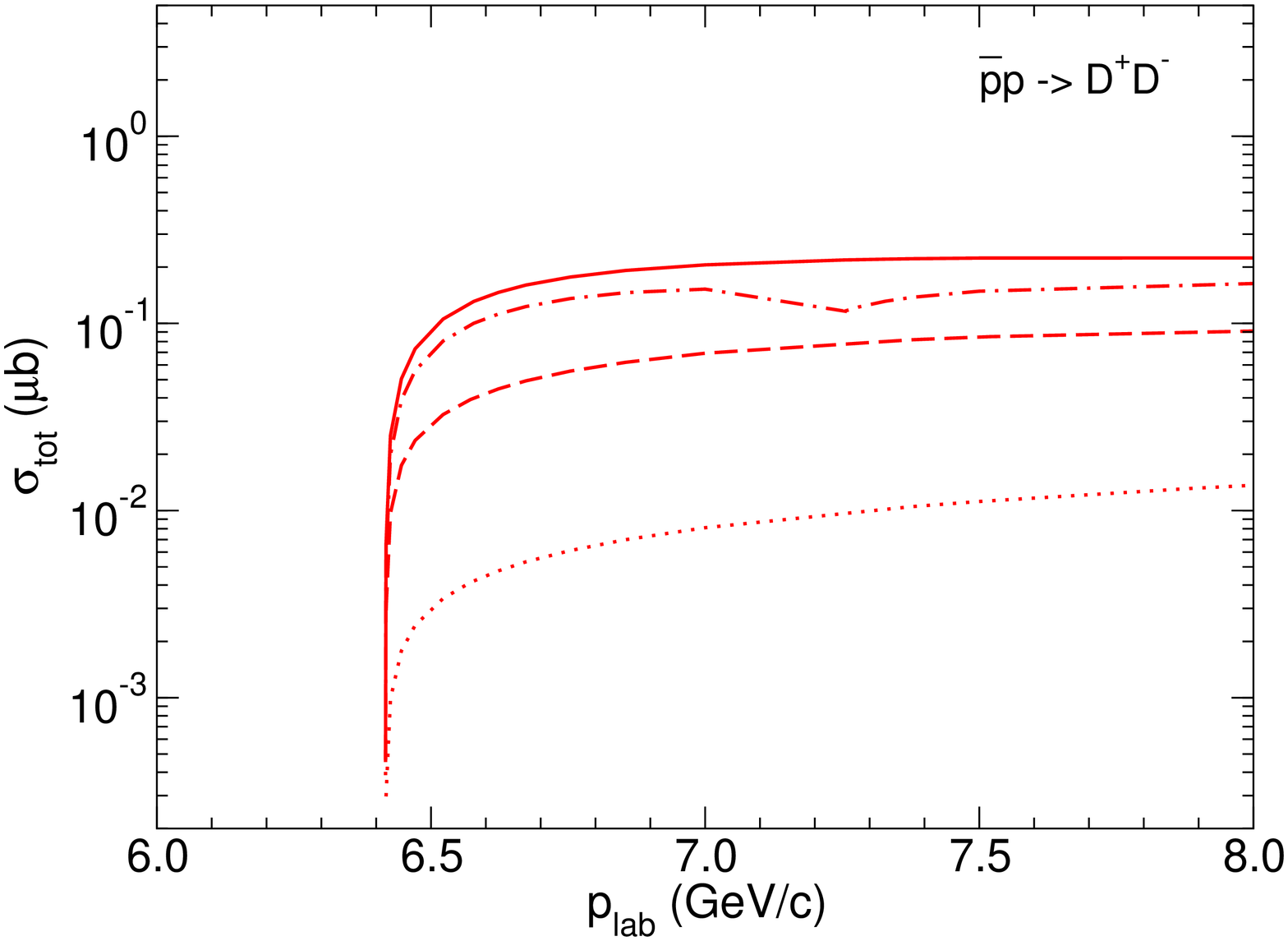}
\caption{Total reaction cross sections for $\ppbar \to \ddbar$ as a function of $p_{lab}$. 
Effects of the final state interaction. The dashed lines are results with the $\nnbar$ model
A as ISI, but without FSI. Inclusion of the $\ddbar$ FSI yields the solid curves. Including
in addition the coupling of $\ddbar$ to $\dsds$ leads to the dash-dotted lines. 
Results obtained in Born approximation are indicated by the dotted lines.
}
\label{fit:fs}   
\end{figure}

In any case, we want to emphasize that one should not take
the quantitative results too literally. It is obvious that 
without any constraints from experiments such model 
calculation are necessarily afflicted with sizeable uncertainties 
as is reflected in the results presented in Ref. \cite{Liu10}. 
The difference in the $\bar DD$ cross sections induced by the 
coupling to the $\dsds$ system shown in Fig.~\ref{DDbcr} may 
serve as further illustration with regard to that. 
But the essential point for our purpose here is that the
$\ddbar$ and $\dsds$ interactions incorporate all essential features
one expects from a realistic FSI. Specifically the amplitudes 
are generated by solving a scattering equation, i.e. they fulfill
unitarity requirements, and they include effects from the 
presence of open channels such as $\pi\pi$ and $\kkbar$.

The uncertainties of our predictions for the reactions $\ppbar \to \ddbar$ 
and $\ppbar \to \dsds$ induced by the treatment of the $\ddbar$ and $\dsds$ 
interactions are best estimated by simply switching off the corresponding 
FSI effects, which will be discussed below. 
Note that such a radical approach supersedes variations coming from a  
possible SU(4) breaking in the coupling constants involved in the 
$\ddbar$ and $\dsds$ interactions, discussed in Appendix \ref{app:su4}.

Results for the reaction $\ppbar \to \ddbar$ are displayed in
Fig.~\ref{fit:fs} where now only cross sections based on the $\nnbar$ 
interaction $A$ are presented so that one can distinguish the
various FSI effects more clearly. For the other variants of the
$\nnbar$ interaction the effects are very similar.
The dotted and dashed lines are again the results obtained in
the Born approximation and by taking into account only the initial
$\nnbar$ interaction, respectively. The inclusion of an interaction in 
the $\ddbar$ system (solid lines) yields a noticeable change in the energy 
dependence of the $D^+D^-$ cross section and an enhancement in the case of the 
$D^0\bar D^0$ channel. 
The $\ddbar$ interaction is only strong in the $I=0$ channel, cf. Fig.~\ref{DDbcr},  
and, therefore, the inclusion of FSI effects modifies primarily the corresponding
$I=0$ $\nnbar \to \ddbar$ transition amplitude. 
Since the $D^+D^-$ and $D^0\bar D^0$ production amplitudes are given by
the coherent sum and difference of the $I=0$ and $I=1$ amplitudes (analogous to 
Eq.~(\ref{Iso})), respectively, the $I=0$ amplitude interferes differently with
the one for $I=1$ for the two particle channels and, accordingly, the FSI effects 
are different.
 
Anyway, overall one can say that the changes are moderate, specifically
if one recalls the variations due the ISI. The results do not change
very much anymore when, finally, also the coupling to the $\dsds$ channel 
is introduced (into the $\ddbar$ FSI), see the dash-dotted lines, though
there is a visible appearance of threshold effects from the opening of the
$\dsds$ channel in the $D^+D^-$ as well as in the $D^0\bar D^0$ cross
sections. Note that the coupling to the $\dsds$ channel has a
sizeable influence on the $\ddbar$ scattering cross section, as said before,
see the dash-dotted curve in Fig.~\ref{DDbcr}.  

\begin{figure}
\includegraphics[height=75mm,angle=-90]{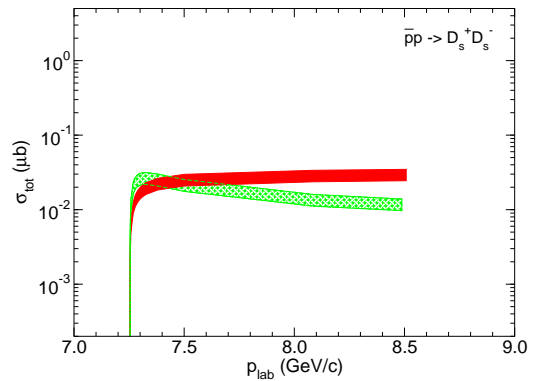}
\caption{Total reaction cross sections for $\ppbar \to \dsds$ as a function of $p_{lab}$, 
based on baryon exchange (shaded band) and the quark model (grid). 
}
\label{fig:ds} 
\end{figure}

Our predictions for the reaction $\ppbar \to \dsds$ can be found in
Fig.~\ref{fig:ds}, where we use the same scale as in the figures with
the $\ppbar \to \ddbar$ results in order to facilitate a comparison. 
Thus, one can see easily that the cross sections for the two
reactions are of comparable magnitude, even though a two-step process 
is required in the former. We should mention that this is not unusual. 
In a calculation of $\bar\Sigma\Sigma$ production, carried out in a similar
framework by our group many years ago \cite{Haiden1993} it was 
found that the cross sections for $\ppbar \to \overline{\Sigma^+}\Sigma^+$ 
and $\ppbar \to \overline{\Sigma^-}\Sigma^-$ were of comparable magnitude. 
Also here the latter reaction requires (at least) a two-step process. 
Indeed, in that case an experiment performed several years later 
\cite{Barnes97} confirmed that the $\overline{\Sigma^-}\Sigma^-$ production 
cross section is not suppressed at all. 

\begin{figure}
\begin{center}
\includegraphics[height=130mm]{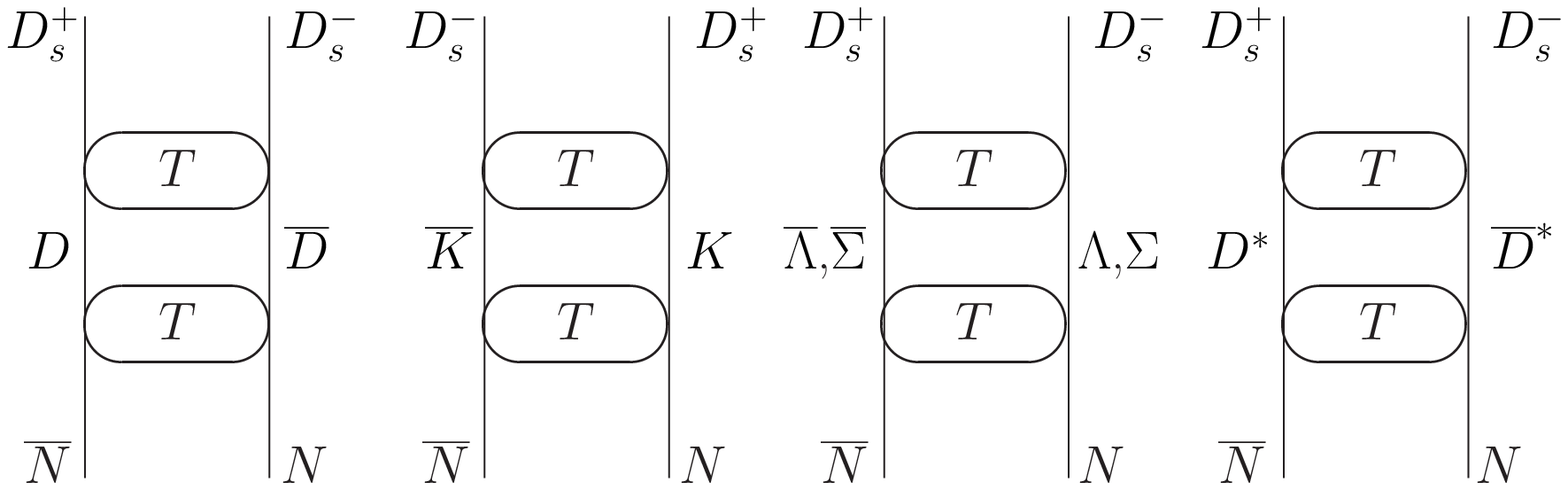}
\vskip -9.0cm 
\caption{Two-step processes that contribute to the reaction 
$\nnbar \to \dsds$. $T$ stands for the corresponding transition 
amplitudes. The two mechanisms on the left side are
included in the present study. 
}
\label{Diads}
\end{center}
\end{figure}

With inclusion of the FSI the amplitudes for $\ppbar \to \ddbar$ and 
$\ppbar \to \dsds$ are given by        
\begin{eqnarray}
\nonumber
T^{\ddbar,\nnbar} &=& (T^{\ddbar,\ddbar} G^{\ddbar} + 1) \\  
                  & & V^{\ddbar,\nnbar} (1 + G^{\nnbar} T^{\nnbar,\nnbar}) \ , \\ 
\nonumber
T^{\dsds,\nnbar} &=& T^{\dsds,\ddbar} G^{\ddbar} \\
\nonumber
                 & & V^{\ddbar,\nnbar} (1 + G^{\nnbar} T^{\nnbar,\nnbar}) \\   
\nonumber
                 &+& T^{\dsds,\kkbar} G^{\kkbar} \\  
                 & & V^{\kkbar,\nnbar} (1 + G^{\nnbar} T^{\nnbar,\nnbar}) \ . 
\label{DWBA1}
\end{eqnarray}

The coupled-channel formalism employed in our calculation implies that 
contributions from the two-step processes $\ppbar \to \ddbar \to \dsds$
and $\ppbar \to \kkbar \to \dsds$ are included (though it turned out
that the latter one is negligibly small). In principle, there are many other 
two-step processes that lead likewise to a final $\dsds$ state. Two examples are 
indicated by the diagrams on the right-hand side of Fig.~\ref{Diads}. These are 
ignored in the present study but, of course, could affect the cross section.
Nevertheless, we expect that the coupling to the $D^*\bar D^*$ 
channel should not change the ($\ddbar$ and $\dsds$) cross sections too dramatically, 
at least for energies below the $D^*\bar D^*$ threshold, based on what we saw in case 
of the $\ddbar$ results and the coupling to $\dsds$ discussed above. 
There is another channel with open charm, namely $D^*\bar D$--$D\bar D^*$, 
for which the threshold is between the ones for $\ddbar$ and $\dsds$. 
Fortunately, it contributes to different parity and total-angular-momentum states 
so that there is no coupling to the $\ddbar$ and $\dsds$ systems.
The thresholds of other possible intermediate states that lead to a final 
$\dsds$ system, like $\ppbar \to \lbarl \to \dsds$, depicted also in Fig.~\ref{Diads}, 
or $\ppbar \to \lcbarlc \to \dsds$ say, are all far away from those of 
$\ddbar$ and $\dsds$ and, therefore, the corresponding two-step processes should 
be less important. 

In any case, it is clear that the predicted $\dsds$ cross section is 
afflicted with larger uncertainties than the one for $\ddbar$. Still, we 
believe that the largest uncertainties come from the form factors in the 
transition potentials and from the $\dsds$ interaction itself. 
Thus, as before in the $\ddbar$ case, we reduced the 
cutoff mass in the vertex form factor of the $\ppbar \to \ddbar$ transition potential 
from 3.5 to 3~GeV, and we also switched off the direct $\dsds$ interaction in order 
to see its effect on the cross section. Both scenarios led to a reduction of the
$\dsds$ yield by a factor of around 4-5. 

Finally, with the aim to shed light on the uncertainties from a different
perspective, we derived also the meson-meson $\kkbar \rightarrow D^+_s D^-_s$ and 
$\ddbar \rightarrow D^+_s D^-_s$ transition amplitudes from the quark-pair 
annihilation-creation processes $\bar{s}s \rightarrow \bar{c}c$ and $\bar{u}u 
\rightarrow \bar{s}s$, respectively; their explicit expressions are summarized 
in Appendix~\ref{app:QM}. 
In Fig.~\ref{mm_qm} we present exemplary results for the cross section in 
the transition reaction $D\bar{D} \rightarrow \dsds$. 
One sees that the prediction obtained from the quark model is on average a 
factor of~4 smaller than the one from the meson-exchange model. 

A direct calculation of $\ppbar \rightarrow \dsds$ based on quark-model 
transition potentials yields a vanishing cross section. This
is so because the transition $\ppbar \to \ddbar$ can only occur
for isospin $I=1$ in the quark model, as said above, while the transition 
$\ddbar \to \dsds$ can take place only in $I=0$ since $\dsds$ is a $I=0$ system. 
Therefore, in the application to the $\ppbar \rightarrow \dsds$ 
we use (somewhat inconsistently) baryon-exchange amplitudes for
$\ppbar \to \ddbar$ in order to estimate the effect of the quark model 
meson-meson transitions on the $\ppbar \to \dsds$ reaction. 
Corresponding results are included in Fig.~\ref{fig:ds}. 
Clearly, both the quark and the 
meson-exchange models yield predictions of comparable magnitude. 
This might be somewhat surprising in view of the cross sections shown 
in Fig.~\ref{mm_qm}. However, one has to keep in mind that the latter
is determined by the on-shell $\ddbar \to \dsds$ $T$-matrix while 
the $\ppbar \to \dsds$ reaction involves this amplitude off-shell,
see Eq.~(\ref{DWBA1}), and here the ones based on the quark and the
meson-exchange models are obviously of similar magnitude. 

\begin{figure}[h]
\includegraphics[height=75mm,angle=-90]{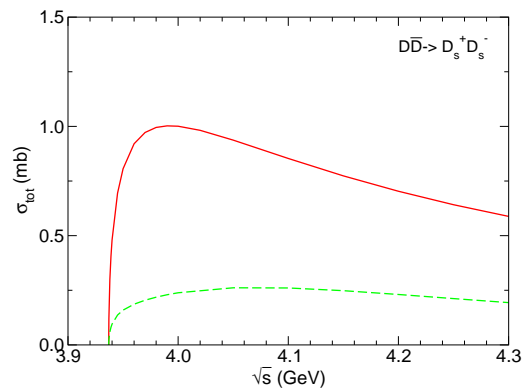}
\caption{Cross section for $D\bar{D} \rightarrow D^+_s D^-_s $ as a 
function of $\sqrt s$ calculated from the meson-exchange model (solid line)
and the quark model (dashed line). 
}
\label{mm_qm}   
\end{figure}

\section{Summary}
\label{sec:5}

We have presented predictions for the reactions $\ppbar \to \ddbar$ and
$\ppbar \to \dsds$ based on a model calculation performed within the 
baryon-exchange picture in close analogy to the J\"ulich analysis of
the reaction $\ppbar \to \kkbar$ \cite{Hippchen2,Mull}, 
connecting those processes via $SU(4)$ symmetry. 
 
Effects of the interaction in the inital $\ppbar$ channel which play a 
crucial role for quantitative predictions are taken into account. 
Furthermore, the J\"ulich $\pi\pi -\kkbar$ model \cite{Janssen95}
was extended to higher energies by including also the $\ddbar$ and
$\dsds$ channels so that even effects from the final-state 
interaction could be investigated. In particular, the coupling between
the $\ddbar$ and $\dsds$ systems, facilitated by the FSI, allows us to 
obtain predictions for $\ppbar \to \dsds$, i.e. for the production
of charmed strange mesons in $\nnbar$ collisions, which is only possible 
via a two-step process.
 
The cross sections for $\ppbar \to \ddbar$ were found to be in the order 
of $10^{-2}$ -- $10^{-1}$ $\mu b$ and they turned out to be comparable 
to those predicted by other model calculations in the literature.
The cross section for a $\dsds$ pair is found to 
be roughly of the same order of magnitude, despite of the fact that its 
production in $\bar pp$ scattering requires a two-step process. 

As before in our study of the reaction $\ppbar \to \lcbarlc$ \cite{HK10}
we investigated an alternative mechanism for the charm production. 
This was done in the form of a $\ppbar \to \ddbar$ transition potential derived 
in a constituent quark model where two (up or down) quark pairs are
annihilated and a charmed quark pair is created. 
It turned out that the $\ppbar \to \ddbar$ cross sections predicted by 
the mechanism based on the quark picture are essentially of the same order 
of magnitude as those that we obtained from baryon exchange. 

Our results suggest that the reactions $\ppbar \to \ddbar$ and
$\ppbar \to \dsds$ take place predominantly in the $s$-wave, at
least for excess energies below 100 MeV, say. 
But we should mention that there is a well-established $p$-wave
resonance, the $\psi(3770)$ ($J^{PC}=1^{--}$) which is seen as 
a pronounced structure in $e^+ e^- \to \ddbar$ \cite{Ablikim,Anashin},
for example, and which decays almost exclusively (i.e. to 93$^{+8}_{-9}$ \%)
into $\ddbar$ \cite{PDG}. This resonance is located at only around 35~MeV 
above the $\ddbar$ threshold. 
We did not include it in the present study because at the moment the 
strength of the coupling of the $\psi(3770)$ to $\ppbar$ is not that well 
known \cite{Ablikim2}. 
But its impact should be definitely explored in any more refined 
studies of $\ppbar \to \ddbar$ in the future. Evidently, it 
would be also interesting to examine the energy range in question 
in pertinent experiments, which could be performed at FAIR, in order 
to see whether there is a signal of this resonance. 

\section*{Acknowledgements}
The work of G.K. was partially financed by CNPq and FAPESP (Brazilian
agencies). The work was also partially supported by the EU Integrated 
Infrastructure Initiative HadronPhysics3 (Contract no. 283286).

\appendix

\section{The interaction Lagrangians}
\label{app:lags}

Here we list the specific interaction Lagrangians which are used to
derive the interactions. 
The baryon-baryon-meson couplings that enter the $\nnbar \to M_1M_2$
transition potentials are given by
\bea
\nonumber
{\mathcal L}_{B'BP} &=& \frac{f_{B'BP}}{m_P} \bar \Psi_{B'}(x) \gamma^5 \gamma^\nu  
\Psi_B(x) \partial_\nu \Phi_P(x) + H.c. \ , \\
{\mathcal L}_{B'\tilde BP} &=& \frac{f_{B'\tilde B P}}{m_P} \bar \Psi_{B'}(x) \Psi_{\tilde B}^\mu(x) 
\partial_\mu \Phi_P (x) + H.c. \ . 
\label{Lag1} 
\eea
In Eq.~(\ref{Lag1}) $\Psi_{B}$ ($B=N, \Lambda, \Sigma, \Lambda_c, \Sigma_c)$ 
are the octet (spin-1/2) baryon field operators 
and $\Psi_{\tilde B}^\mu$ ($\tilde B=\Delta, \Sigma^*(1385), \Sigma_c^*(2520)$) 
the decuplet (spin-3/2) field operators,
while $\Phi_P$ is the field operator for pseudoscalar mesons. 
Explicit expression for the resulting transition potentials can be found in
the Appendix A of Ref.~\cite{Hippchen1}. 

The employed three-meson couplings for the various $M_1M_2 \to M_3M_4$ potentials
and transitions are
\begin{eqnarray}
\nonumber
{\mathcal L}_{PPS} &=& \frac{g_{PPS}}{m_P} \ \partial^\mu\Phi_P(x) \partial_\mu\Phi_P(x) \Phi_S(x) \ , \\
\nonumber
{\mathcal L}_{PPV} &=& g_{PPV} \ \Phi_P(x) \partial_{\mu} \Phi_P(x) \Phi_V^{\mu}(x) \ , \\
{\mathcal L}_{PPT} &=& g_{PPT} \ \partial_\mu\Phi_P(x) \partial_{\nu} \Phi_P(x) \Phi_T^{\mu\nu}(x) \ , 
\label{Lag2}
\end{eqnarray}
for the coupling of a scalar ($S$), vector ($V$), or tensor ($T$) meson to pseudoscalar
mesons. Expression for the transition potentials in the meson-meson sector can be found 
in the Appendix of Ref.~\cite{Janssen95}. 
Note that in the equations above only the space-spin part is 
given. There is also an isospin dependence that has to be taken into account in
the actual calculation. 
The SU(4) flavour structure leads to the characteristic relations between the coupling constants.
For the vertices involving baryons they are given by 
\begin{eqnarray}
f_{\Lambda_c^+ N D} &=& f_{\Lambda N K} = -\frac{1}{\sqrt{3}}(1+2{\alpha}) f_{NN\pi}, 
\nonumber \\
f_{\Sigma_c N D } &=& f_{\Sigma N K} =(1-2{\alpha}) f_{NN\pi}, 
\nonumber \\
f_{N\Sigma_c^* D} &=& f_{N\Sigma^* K} = -\frac{1}{\sqrt{6}} f_{N\Delta \pi} , 
\label{SiCou2}
\end{eqnarray}
with $\alpha$ the $F/(F+D)$ ratio. 
The coupling constants for the meson-meson interaction are discussed in detail below. 

\section{Quark model expressions}
\label{app:QM}

The result for the $\ppbar \rightarrow \ddbar$ transition potential presented
in Eq.~(\ref{VppQM}) is obtained from the matrix element $\langle \Psi_f|V_A|\Psi_i\rangle$
provided in Eq.~(A.2) of Ref.~\cite{KohnoM}, with the relative $\ppbar$ wave function 
$\varphi_E({\bf r})$ given by a plane wave with momentum ${\bf p}$, and after summing 
over spin-isospin indices. The explicit expressions of the form factors $h_1(t) = 
h_1({\bf p},{\bf p}')$ and $h_2(t) = h_2({\bf p},{\bf p}')$ that appear in Eq.~(\ref{VppQM}) 
are given by 
\bea
h_1({\bf p},{\bf p}') &=& - \frac{b^2_N}{3b^2_N + 2 b^2_M}\, h({\bf p},{\bf p}') ,
\label{h1}
\\[0.2true cm]
h_2({\bf p},{\bf p}') &=&  \left[1 - 2\beta \frac{b^2_M}{(3 b^2_N + 2 b^2_M)}\right] 
\, h({\bf p},{\bf p}') ,
\label{h2}
\eea
with $\beta = m_q/(m_q+m_h)$, $m_q = (m_u, m_d)$, $m_h = (m_c, m_s)$, and 
\bea
h({\bf p},{\bf p}') &=& 24 \, \lambda \, C_A \frac{4\pi}{Q} 
\left(\frac{\alpha_A}{m^2_G}\right)^2  
v({\bf p},{\bf p}') \nonumber \\
&\times& 4\pi  \int^\infty_0 dz \, z^3 \, j_1(Qz) \, f(z) ,
\label{h}
\eea
where $C_A = 4/27$ comes from summing over color indices, $\lambda = 1$ for 
$\ppbar \rightarrow \bar D^0 D^0\, (K^+ K^-)$,  $\lambda = - 1$ for $\ppbar \rightarrow 
\bar D^+ D^-\, (\bar K^0 K^0)$,
\beq
{\bf Q} = \left[1 - 2\beta \frac{b^2_M}{(3 b^2_N + 2 b^2_M)}\right] \, {\bf p}' 
- \frac{b^2_N}{3b^2_N + 2 b^2_M}\,{\bf p} , 
\label{Q}
\eeq
and
\bea
v({\bf p},{\bf p}') &=& \left[\frac{12b^2_M}{(3b^2_N + 5 b^2_M)(3 b^2_N + 2 b^2_M)}
\right]^{3/2}  \nonumber \\[0.2true cm]
&\times & \exp\left[-\frac{b^2_N b^2_M}{9b^2_N + 6 b^2_M}\, 
({\bf p} - 3 \beta {\bf p}')^2\right] ,
\label{vppm}
\\[0.5true cm]
f(z) &=& 
\frac{1}{z^3} \left(1 + \mu z\right) \nonumber \\
&\times & \exp \left[  - \mu z
- \left(\frac{2b^2_N + b^2_M}{3b^2_N + 2 b^2_M}\right) z^2 \right] .
\label{fz}
\eea 
Here, $\mu = m_h$ comes from the (static) heavy quark propagator - see Fig.~\ref{fig:qqQQ}, 
and $b_N$ and $b_M$ are the Gaussian widths of the nucleon and meson wave 
functions, related to their rms radii by $b_N = \langle r^2_N \rangle^{1/2}$ 
and $b_M = \sqrt{8/3} \, \langle r^2_M \rangle^{1/2}$. Also, in Eq.~(\ref{h}), the 
factor $(\alpha_A/m^2_G)$ indicates the strengths of light-quark pair annihilation
and heavy-quark pair creation. 

For the $\kkbar \rightarrow D^+_s D^-_s$ transition potential, the explicit expression 
is 
\begin{eqnarray}
&& V^{\kkbar \rightarrow D^+_sD^-_s}_Q({\bf p},{\bf p}') =  4\pi \frac {\alpha_A}{m^2_G}
\,C_{KD_s} \,  \left(\frac{2 \, b_{K} \, b_{D_s}}{b^2_{K} + b^2_{D_s}}\right)^3 
\nonumber \\[0.2true cm]
&&\hspace{1.5cm}\times \, \exp \left[ - \frac{1}{4} \frac{b^2_K b^2_{D_s}}{b^2_K + b^2_{D_s}}
\left(m {\bf p} - M {\bf p}'\right)^2\right],
\end{eqnarray} 
where ${\bf p}$ and ${\bf p}'$ are the initial and final state c.m. momenta, 
$C_{KD_S} = 4/9$ comes from color, 
\begin{equation}
m = \frac{2 m_s}{m_s + m_q}, \hspace{1.0cm} M = \frac{2m_s}{m_c + m_s},
\label{m&M-KDs}
\end{equation}
and the $b$'s are Gaussian size parameters, related to the meson r.m.s. radii as 
above. For the reaction $\ddbar \rightarrow D^+_s D^-_s$, the color factor is the same, 
$b_K \rightarrow b_D$, and the masses $m$ and $M$ are replaced by
\begin{equation}
m = \frac{2 m_c}{m_c + m_q}, \hspace{1.0cm} M = \frac{2m_c}{m_c + m_s}.
\label{m&M-DDs}
\end{equation}

The values of parameters we use are standard, namely: the constituent quark masses are 
taken to be $m_u = m_d = 330$~MeV, $m_s = 550$~MeV, $m_c = 1600$~MeV, and the meson
Gaussian size parameters are such that $\langle r^2_M \rangle^{1/2} = 0.4$~fm
and $\langle r^2_N \rangle^{1/2} = 0.55$~fm. 
Regarding the effective coupling strength we use the value 
$\alpha_A/m^2_G = 0.12$~fm$^2$ which we fixed from a fit to 
the $\ppbar \to K^-K^+$ cross section, i.e. in a reaction where a pair of
strange quarks is produced. This is very close to the original value of 
Kohno and Weise~\cite{KohnoM} who adopted the value $\alpha_A/m^2_G = 0.15$~fm$^2$ 
in their calculation of the same reaction. 
As already noted in our study of the reaction $\ppbar \rightarrow \lcbarlc$ 
\cite{HK10}, this effective coupling depends implicitly on the 
effective gluon propagator, i.e. on the square of the energy transfer from initial 
to final quark pair. Heuristically, this energy transfer corresponds
roughly to the masses of the produced constituent quarks,
i.e. $m_G \approx 2 m_q$. Thus, we assume the effective coupling strength
for charm production to be reduced by the ratio of the constituent quark masses 
of the strange and the charmed quark squared, $(m_s/m_c)^2 \approx (550/1600)^2 
\approx 1/9$ as compared to the one used for strangeness production. 

\section{SU(4) relations for the meson-meson interaction}
\label{app:su4}

The general form of the SU(4) invariant Lagrangian~is
\begin{widetext}
\begin{eqnarray}
\nonumber
{\mathcal L}_{MMM} & = & g_{\{15\}}\left[-\alpha Tr([M_{\{15\}},M_{\{15\}}]M_{\{15\}}) +
                  (1-\alpha) Tr(\{M_{\{15\}},M_{\{15\}}\}M_{\{15\}}) \right] \\ \nonumber
& + & g_{\{15\}\{15\}\{1\}} (1-\alpha) Tr(\{M_{\{15\}},M_{\{15\}}\}M_{\{1\}}) +
                g_{\{15\}\{1\}\{15\}}(1-\alpha) Tr(\{M_{\{15\}},M_{\{1\}}\}M_{\{15\}}) \\
& + & g_{\{1\}}(1-\alpha) Tr(\{M_{\{1\}},M_{\{1\}}\}M_{\{1\}}) \ ,
\label{lsu4}
\end{eqnarray}
\end{widetext}
where $M_{\{15\}}$ ($M_{\{1\}}$) stands for the SU(4) meson--15-plet (-singlet) matrix. 
For pseudo-scalar ($P$) and vector ($V$)
mesons $M_{\{15\}}$ is a $4\times4$ matrix of the form
\begin{widetext}
\begin{eqnarray} P&=&
\left (
\begin{array}{cccc}
\frac{\pi^0}{\sqrt 2}+\frac{\eta}{\sqrt 6}+\frac{\eta_c}{\sqrt{12}}
& \pi^+ & K^+ & {\bar D}^0 \\
\pi^- & -\frac{\pi^0}{\sqrt 2}+\frac{\eta}{\sqrt 6}+\frac{\eta_c}{\sqrt{12}}
& K^0 & D^- \\
K^- & {\bar K}^0 & -\sqrt {\frac{2}{3}}\eta+\frac{\eta_c}{\sqrt{12}}
& D_s^- \\
D^0 & D^+ & D_s^+ & -\frac{3\eta_c}{\sqrt{12}}
\end{array}
\right ) \;, \nonumber \\[2ex]
V&=&
\left (
\begin{array}{cccc}
\frac{\rho^0}{\sqrt 2}+\frac{\omega_8}{\sqrt 6}+\frac{\omega_{15}}{\sqrt {12}}
& \rho^+ & K^{*+} & {\bar D}^{*0} \\
\rho^- & -\frac{\rho^0}{\sqrt 2}+\frac{\omega_8}{\sqrt 6}+\frac{\omega_{15}}
{\sqrt {12}} & K^{*0} & D^{*-} \\
K^{*-} & {\bar K}^{*0} & -\sqrt {\frac{2}{3}}\omega_8+\frac{\omega_{15}}{\sqrt {12}}
& D_s^{*-} \\
D^{*0} & D^{*+} & D_s^{*+} & -\frac{3\omega_{15}}{\sqrt {12}}
\end{array}
\right ) \;. \nonumber
\end{eqnarray}
\end{widetext}
 
For the construction of the $\ddbar$ and $\dsds$ interactions and the
transition potentials to the $\pi\pi$, $\pi\eta$, and $\kkbar$ channels we 
need three-meson vertices involving charmed mesons of the kind $PPV$.
The $PPV$ vertices involve only $F$-type coupling ($\alpha = 1$)
if we require charge conjugation invariance and, therefore, 
in this case there is no singlet coupling, cf. Eq.~(\ref{lsu4}). 

Based on the assumed SU(4) symmetry all relevant three-meson coupling constants
can be derived from the empirically known $\pi\pi\rho$ coupling. In the
J\"ulich model \cite{Janssen95} the value $g_{\pi\pi\rho} = 6.04$ is used. 
The coupling constants of the other vertices that follow from this value
are listed in Table \ref{coup}.

{ \renewcommand{\arraystretch}{1.4} 
\begin{table}
\caption{Vertex parameters for $t$-channel exchanges.  Relations
between coupling constants are obtained using SU(4) and ideal 
mixing between the octet and singlet.}
\label{coup}
\begin{tabular}{ccr}
\hline 
\hline
Vertex & $g$ & \multicolumn{1}{c}{$\Lambda$ [MeV]} \\ 
\hline
$\pi\pi\rho$ & 6.04 & 1355 \\ 
$\pi KK^*$ & $g_{\pi KK^*} = g_{\pi
\overline{K}\overline{K^*}} = -\frac{1}{2}g_{\pi\pi\rho}$ & 1900 \\
$KK\rho$ & $g_{KK\rho} = g_{\overline{K}\overline{K}\rho} =
\frac{1}{2}g_{\pi\pi\rho}$ & 1850 \\ 
$KK\omega$ & $g_{KK\omega} =
-g_{\overline{K}\overline{K}\omega} = \frac{1}{2}g_{\pi\pi\rho}$ & 2800 \\ 
$KK\phi$ & $g_{KK\phi} = -g_{\overline{K}\overline{K}\phi} =
\frac{1}{\sqrt{2}}g_{\pi\pi\rho}$ & 2800 \\ 
$\eta KK^*$ & $g_{\eta KK^*} = -g_{\eta \overline{K}\overline{K^*}} =
-\frac{\sqrt{3}}{2}g_{\pi\pi\rho}$ & 3290 \\
\hline
$\pi DD^*$ & 
$g_{\pi \overline{D}\overline{D^*}} = g_{\pi DD^*} = 
-\frac{1}{2}g_{\pi\pi\rho}$ & 3100 \\
$DKD_s^*$ &
$g_{\bar D K D_s^{*-}} = 
-g_{D \bar K D_s^{*+}} =-\frac{1}{\sqrt{2}}g_{\pi\pi\rho}$ & 3100 \\
$DD\rho$ &
$g_{\overline{D}\overline{D}\rho} = g_{DD\rho} = 
\frac{1}{2}g_{\pi\pi\rho}$ & 1850 \\ 
$DD\omega$ &
$g_{\overline{D}\overline{D}\omega} = -g_{DD\omega} =
\frac{1}{2}g_{\pi\pi\rho}$ & 2800 \\ 
$DD\psi$ &
$g_{\overline{D}\overline{D}\psi} = -g_{DD\psi} = 
-\frac{1}{\sqrt{2}}g_{\pi\pi\rho}$ & 4000 \\ 
\hline
$D_s K D^*$ &
$g_{D_s^+ K D^*} = 
-g_{D_s^- \bar K \bar D^*}= \frac{1}{\sqrt{2}}g_{\pi\pi\rho}$ & 3100 \\
$DD_s K^*$ &
$g_{\bar D D_s^- K^*} = 
-g_{D  D_s^+ K^*}= \frac{1}{\sqrt{2}}g_{\pi\pi\rho}$ & 1900 \\
$D_sD_s\phi$ &
$g_{D_s^-D_s^-\phi} = -g_{D_s^+D_s^+\phi} =
-\frac{1}{\sqrt{2}}g_{\pi\pi\rho}$ & 2800 \\ 
$D_sD_s\psi$ &
$g_{D_s^-D_s^-\psi} = -g_{D_s^+D_s^+\psi} = 
-\frac{1}{\sqrt{2}}g_{\pi\pi\rho}$ & 4000 \\ 
\hline
\end{tabular}
\end{table}
}

Let us make some more comments about the coupling constants at the
three-meson vertices, specifically with regard to the imposed 
ideal mixing between the octet and singlet. 
$SU(4)$ symmetry implies the following for the 
vector meson coupling constants relevant for our study:
\begin{eqnarray}
\nonumber
g_{KK\omega_8}&=&\sqrt{3} g_{KK\rho } = \frac{\sqrt{3}}{2}\, g_{\pi\pi\rho} , \ \
g_{KK\omega_{15}}=0 , \\
\nonumber
g_{\bar D \bar D\omega_8}&=&\sqrt{\frac{1}{3}} g_{KK\rho} , \ \
g_{\bar D \bar D\omega_{15}}=\sqrt{\frac{8}{3}} g_{KK\rho}, \\
\nonumber
g_{D_s^- D_s^-\omega_8}&=&-\frac{2}{\sqrt{3}} g_{KK\rho} , \ \
g_{D_s^- D_s^-\omega_{15}}=\sqrt{\frac{8}{3}} g_{KK\rho}, \\
\nonumber
g_{D_s^+ D_s^+\omega_8}&=&\frac{2}{\sqrt{3}} g_{KK\rho} , \ \
g_{D_s^+ D_s^+\omega_{15}}=-\sqrt{\frac{8}{3}} g_{KK\rho} . \\
\label{omega}
\end{eqnarray}

Assuming ideal mixing of the $\omega_{15}$, $\omega_8$ and $\omega_1$ one obtains
for the coupling constants of the physical $\omega$, $\phi$, and $J/\psi$ 
\begin{eqnarray}
\nonumber
g_{\bar D\bar D\omega}&=&\sqrt{\frac{1}{2}}g_{\bar D\bar D\omega_1}+
\sqrt{\frac{1}{3}}g_{\bar D\bar D\omega_8} +
\sqrt{\frac{1}{6}}g_{\bar D\bar D\omega_{15}}, \\
\nonumber
g_{\bar D\bar D\phi}&=&-\sqrt{\frac{1}{4}}g_{\bar D\bar D\omega_1}+
\sqrt{\frac{2}{3}}g_{\bar D\bar D\omega_8} -
\sqrt{\frac{1}{12}}g_{\bar D\bar D\omega_{15}}, \\
g_{\bar D\bar D\psi}&=& \sqrt{\frac{1}{4}}g_{\bar D\bar D\omega_1}-
\sqrt{\frac{3}{4}}g_{\bar D\bar D\omega_{15}} \ .
\end{eqnarray}
The same relation holds also for the $K$ meson and for the $D_s^-$ and $D_s^+$. 
In case of the $K$ meson
the coupling constant $g_{KK\omega}$ is given by that of $g_{KK\omega_8}$ alone,
since there is no singlet coupling for PPV vertices as mentioned above:
\begin{eqnarray}
g_{KK\omega}&=&\sqrt{\frac{1}{3}} g_{KK\omega_8} = g_{KK\rho} \ .
\label{omegaK}
\end{eqnarray}
In case of the $D$ meson the coupling constant is given by
\begin{eqnarray}
\nonumber
g_{\bar D\bar D\omega}&=&\sqrt{\frac{1}{3}}g_{\bar D\bar D\omega_8} +
\sqrt{\frac{1}{6}}g_{\bar D\bar D\omega_{15}}
= g_{KK\rho} \\
g_{\bar D\bar D\psi}&=-&\sqrt{\frac{3}{4}}g_{\bar D\bar D\omega_{15}}
= -\sqrt{2}\, g_{KK\rho} \ ,
\label{omegaD}
\end{eqnarray}
and for the $D_s$ meson,
\begin{eqnarray}
\nonumber
g_{D_s^-D_s^-\phi}&=&\sqrt{\frac{2}{3}}g_{D_s^-D_s^-\omega_8} -
\sqrt{\frac{1}{12}}g_{D_s^-D_s^-\omega_{15}} \\
\nonumber
&=& -\sqrt{2}\, g_{KK\rho} \\ 
g_{D_s^-D_s^-\psi}&=-&\sqrt{\frac{3}{4}}g_{D_s^-D_s^-\omega_{15}} 
= -\sqrt{2}\, g_{KK\rho} \ .
\label{omegaDs}
\end{eqnarray}

The J\"ulich $\pi\pi-\kkbar$ potential contains also 
vertex form factors $F$ that are meant to take into account the 
extended hadron structure and are parametrized in the conventional 
monopole or dipole form \cite{Lohse90,Janssen95}. In the present 
extension to the $\ddbar$ and $D_s^+D_s^-$ systems
the cut-off masses appearing in those form factors for the various
three-meson vertices are mostly taken over from Ref.~\cite{Janssen95}. 
In particular, for vertices involving vector mesons without charm
($\rho$, $\omega$, $\phi$, $K^*$), we make the assumption that 
$F_{D D V}({\bf q}_V^{\, 2}) \approx F_{KKV}({\bf q}_V^{\, 2})$  
and/or 
$F_{D_sD_s V}({\bf q}_V^{\, 2}) \approx F_{KKV}({\bf q}_V^{\, 2})$,
i.e. we use the same cutoff masses for the same exchange particles --
a prescription that is guided by the notion that those form factors parametrize 
predominantly the off-mass-shell behavior of the exchanged particles. 
For the additional vertices that concern the exchange of a $D^*$(2009) or
$D_s^*$(2112) or of a $\psi$(3096) we adopt cutoff masses that are about 
1~GeV larger than the mass of the exchange particle. 
A compilation of the cutoff masses employed at the various three-meson
vertices is provided in Table~\ref{coup}. 

In our model calculation we use $PPV$ coupling constants that
are determined fully by SU(4) symmetry. In our opinion, the difference 
of those values to the ones deduced from available experimental information 
is not very large and, thus, does not really warrant a departure from SU(4) 
at present. But let us review the situation briefly here. 
The $DD\rho$ coupling constant was determined in Refs.~\cite{Mat98,Lin00a} 
based on the vector dominance model and found to be
$g_{DD\rho} = 2.52-2.8$. This value, which was subsequently adopted in several
investigations~\cite{Lin00,Lin01,Liu02}, is only marginally smaller than
the one which follows from assuming SU(4) symmetry. The same is true for the
$DD\omega$ coupling constant, found to be $g_{DD\omega} = -2.84$ in 
Ref. \cite{Lin00a}, likewise derived within the vector dominance model. 
In Ref. \cite{Liu02} the value $g_{\pi D D^*} = 5.56$ is cited, derived
from the measured decay width of the $D^*$ meson. Here the corresponding
SU(4) coupling constant is roughly a factor 2 smaller. Deviations from the SU(4) 
symmetry are also discussed in Refs.~\cite{ElBennich11,{Krein:2012lra},{bracco},{can}}. 


%

\end{document}